\documentclass[twocolumn]{aastex62}

\usepackage{amssymb, amsmath, amsthm}
\usepackage{graphicx}
\usepackage{multirow}
\hypersetup{linkcolor=red,citecolor=blue,filecolor=cyan,urlcolor=blue}

\begin{document}

\title{The JCMT BISTRO-2 Survey: Magnetic Fields of the Massive DR21 Filament}


\author[0000-0001-8516-2532]{Tao-Chung Ching} 
\email{chingtaochung@gmail.com}
\affiliation{Research Center for Intelligent Computing Platforms, Zhejiang Lab, Hangzhou 311100, People's Republic of China}                                                                                                                                                                                                        
\affiliation{National Astronomical Observatories, Chinese Academy of Sciences, A20 Datun Road, Chaoyang District, Beijing 100012, People's Republic of China}
\affiliation{Jansky Fellow, National Radio Astronomy Observatory, 1003 Lopezville Road, Socorro, NM 87801, USA}

\author[0000-0002-5093-5088]{Keping Qiu}
\affiliation{School of Astronomy and Space Science, Nanjing University, 163 Xianlin Avenue, Nanjing 210023, People's Republic of China}
\affiliation{Key Laboratory of Modern Astronomy and Astrophysics (Nanjing University), Ministry of Education, Nanjing 210023, People's Republic of China}

\author[0000-0003-3010-7661]{Di Li}
\affiliation{National Astronomical Observatories, Chinese Academy of Sciences, A20 Datun Road, Chaoyang District, Beijing 100012, People's Republic of China}
\affiliation{Department of Astronomy, University of Chinese Academy of Sciences, Beijing 100049, People's Republic of China}
\affiliation{NAOC-UKZN Computational Astrophysics Centre, University of KwaZulu-Natal, Durban 4000, South Africa}

\author[0000-0003-4659-1742]{Zhiyuan Ren}                                                                                                                                                                                                         
\affiliation{National Astronomical Observatories, Chinese Academy of Sciences, A20 Datun Road, Chaoyang District, Beijing 100012, People's Republic of China}

\author[0000-0001-5522-486X]{Shih-Ping Lai}
\affiliation{Institute of Astronomy and Department of Physics, National Tsing Hua University, Hsinchu 30013, Taiwan}
\affiliation{Academia Sinica Institute of Astronomy and Astrophysics, No.1, Sec. 4., Roosevelt Road, Taipei 10617, Taiwan}

\author[0000-0001-6524-2447]{David Berry}
\affiliation{East Asian Observatory, 660 N. A'oh\={o}k\={u} Place, University Park, Hilo, HI 96720, USA}

\author[0000-0002-8557-3582]{Kate Pattle}
\affiliation{Department of Physics and Astronomy, University College London, Gower Street, London WC1E 6BT, United Kingdom}

\author{Ray Furuya}
\affiliation{Tokushima University, Minami Jousanajima-machi 1-1, Tokushima 770-8502, Japan}
\affiliation{Institute of Liberal Arts and Sciences Tokushima University, Minami Jousanajima-machi 1-1, Tokushima 770-8502, Japan}

\author[0000-0003-1140-2761]{Derek Ward-Thompson}
\affiliation{Jeremiah Horrocks Institute, University of Central Lancashire, Preston PR1 2HE, UK}

\author[0000-0002-6773-459X]{Doug Johnstone}
\affiliation{NRC Herzberg Astronomy and Astrophysics, 5071 West Saanich Road, Victoria, BC V9E 2E7, Canada}
\affiliation{Department of Physics and Astronomy, University of Victoria, Victoria, BC V8W 2Y2, Canada}

% ...

\begin{abstract}
We present 850 $\mu$m dust polarization observations of the massive DR21 filament from the B-fields In STar-forming Region Observations (BISTRO) survey, using the POL-2 polarimeter and the SCUBA-2 camera on the James Clerk Maxwell Telescope. We detect ordered magnetic fields perpendicular to the parsec-scale ridge of the DR21 main filament. In the sub-filaments, the magnetic fields are mainly parallel to the filamentary structures and smoothly connect to the magnetic fields of the main filament. We compare the POL-2 and Planck dust polarization observations to study the magnetic field structures of the DR21 filament on 0.1--10 pc scales. The magnetic fields revealed in the Planck data are well aligned with those of the POL-2 data, indicating a smooth variation of magnetic fields from large to small scales. The plane-of-sky magnetic field strengths derived from angular dispersion functions of dust polarization are 0.6--1.0 mG in the DR21 filament and $\sim$ 0.1 mG in the surrounding ambient gas.
The mass-to-flux ratios are found to be magnetically supercritical in the filament and slightly subcritical to nearly critical in the ambient gas. The alignment between column density structures and magnetic fields changes from random alignment in the low-density ambient gas probed by Planck to mostly perpendicular in the high-density main filament probed by JCMT. The magnetic field structures of the DR21 filament are in agreement with MHD simulations of a strongly magnetized medium, suggesting that magnetic fields play an important role in shaping the DR21 main filament and sub-filaments.
\end{abstract}


\section{Introduction}
Recent observations of thermal continuum from dust and molecular lines from gas have revealed that parsec-scale filaments are ubiquitous structures in molecular clouds \citep{2014Andre}. 
The collapse and fragmentation of gravitationally unstable filaments host the birth of prestellar cores and protostars \citep{2010Molinari, 2011Arzoumanian, 2013Hacar, 2013Palmeirim, 2014FL, 2015Konyves}. 
Further, high-mass star-forming regions are preferentially found in the hubs of filaments, where the longitudinal mass flows along filaments toward the hubs are believed to play a key role in enhancing the density to drive massive star formation  \citep{2010GM, 2011Hill, 2012Hennemann, 2012Liu, 2012Schneider, 2013Peretto, 2018Hacar, 2020Kumar}.

Observations of dust polarization at submillimeter/millimeter wavelengths have been proven to be the most efficient method to trace magnetic fields of molecular clouds \citep{2012Crutcher}, given that the emission of magnetically aligned interstellar dust grains is linearly polarized with the polarization angle perpendicular to the direction of local magnetic field projected on the plane of sky \citep{2007LH,2015Andersson}.
Single-dish dust polarization surveys reveal magnetic field structures within molecular clouds at resolutions from a few arcmins to tens of arcsecs \citep[e.g.][]{2000Dotson, 2010Dotson, 2009Matthews, 2015PlanckXIX}.
Statistical studies of Planck data covering column densities from 10$^{20}$ to 10$^{22}$ cm$^{-2}$ indicate that the low column density structures  in diffuse clouds appear to be parallel to the magnetic fields, while the filamentary structures of molecular clouds with high column densities tend to be perpendicular to the magnetic fields \citep{2016PlanckXXXII,2016PlanckXXXV}. 
Ground-based telescopes that are capable of resolving magnetic fields in molecular clouds show that at parsec scale, the magnetic fields of filaments are usually perpendicular to the main axes of filaments \citep{1998Schleuning, 2006VF, 2014Matthews, 2017Pattle, 2018Liu, 2019Chuss, 2019Fissel, 2019Soam}. 
The parallel alignment between magnetic fields and low-density sub-filaments and the perpendicular alignment between magnetic fields and high-density filaments are also supported by optical and infrared polarization data, indicating that magnetic fields play an important role in filament formation \citep{2008Alves, 2011Sugitani, 2013Palmeirim, 2016Soler, 2016Cox, 2020Wang}.
Observations at a few thousand au resolution toward dense cores within filaments, however, reveal complex magnetic fields that are not simply aligned with the structures of cores \citep{2014Zhang, 2014Koch, 2015Li, 2020Doi, 2021Eswaraiah}, indicating a more complex role of magnetic fields in the formation of dense cores. 
 
To study the role of magnetic fields in the formation of filaments and high-mass star-forming cores, we present 850 $\mu$m dust polarization observations taken using the James Clerk Maxwell Telescope (JCMT) toward the DR21 filament.
The DR21 filament is the densest and most massive region in the Cygnus X complex \citep{2016Schneider, 2019Cao} at a distance of 1.4 kpc \citep{2012Rygl}.
The filament hosts 24 massive dense cores \citep{2007Motte}, including the well-studied massive star-forming regions DR21 and DR21(OH) \citep{1966DR}.
The ridge of the DR21 filament has a length of 4 pc and a total mass of 15,000 M$_\sun$, connected by several sub-filaments with masses between 130 M$_\sun$ and 1400 M$_\sun$ \citep{2012Hennemann}.
Global infall motions of the filament are suggested by molecular line observations, probably triggered by convergence of flows on cloud scales \citep{2010Schneider, 2011Csengeri}.
Embedded clusters of young stellar objects \citep{2007Kumar}, prominent outflows \citep{2007Davis, 2007Motte, 2013Duarte, 2014Duarte, 2018Ching}, and masers \citep{1983BE, 2000Argon, 2005Pestalozzi} are found in the filament, indicating recent high- to intermediate-mass star formation.
The active star formation of the DR21 filament could be driven by both the mass accretion through the sub-filaments and the converging flows of clouds \citep{2010Schneider, 2012Hennemann}. 

The magnetic fields of the DR21 filament have been mapped through single-dish observations of dust polarized emission (100 $\mu$m at 35$\arcsec$ resolution: \citealt{2000Dotson}; 350 $\mu$m at 10$\arcsec$ and 20$\arcsec$ resolutions: \citealt{2009Kirby, 2010Dotson}; 800 $\mu$m at 14$\arcsec$ resolution: \citealt{1994MM, 1999Greaves}; 850 $\mu$m at 14$\arcsec$ resolution: \citealt{2006VF, 2009Matthews}; 1.1 mm at 19$\arcsec$ resolution: \citealt{1999Greaves}; 1.3 mm at 33$\arcsec$ resolution:  \citealt{1999Glenn}), revealing a uniform structure of magnetic fields at parsec scale that is perpendicular to the filament.   
Single-dish observations of CN Zeeman measurements at a resolution of 23$\arcsec$ (0.16 pc) found line-of-sight magnetic field strengths of 0.4--0.7 mG in DR21 (OH) \citep{1999Crutcher, 2008Falgarone}, and interferometric HI Zeeman observations at a resolution of  5\arcsec (0.03 pc) found a line-of-sight magnetic field strength of a few tenths mG toward the compact HII region of the DR21 core \citep{1997Roberts}.
In contrast to the uniform magnetic fields of the filament, interferometric dust polarization observations reveal complex magnetic field structures in the massive dense cores of the filament, suggesting that the magnetic field plays a more important role in the formation of the DR21 filament than in the formation of the cores \citep{2003Lai, 2013Girart, 2017Ching}.
A combined analysis of dust polarization data and molecular line data suggests that the gas dynamics arising from gravitational collapse may be the origin of distortion of the magnetic fields in the cores \citep{2018Ching}.

Our observations toward the DR21 filament are part of the extension of the B-fields In STar-forming Region Observations (BISTRO) survey \citep{2017Ward}. 
The BISTRO-1 survey carried out POL-2 observations from 2016 to 2019 toward nearby star-forming regions of the Gould Belt clouds, including Orion A \citep{2017Pattle, 2021Hwang}, Ophiuchus \citep{2018Kwon, 2018Soam, 2019Liu}, IC 5146 \citep{2019Wang}, Barnard 1 \citep{2019Coude}, NGC 1333 \citep{2020Doi, 2021Doi}, Auriga \citep{2021Ngoc}, Taurus \citep{2021Eswaraiah}, Orion B \citep{2021Lyo}, and Serpens \citep{2022Kwon}, aiming to generate a large sample of polarization maps in a uniform and consistent way to study the role of magnetic fields in star formation at a few thousand au scales.
The BISTRO-1 survey was later extended to the BISTRO-2 program for high-mass star forming regions (M16: \citealt{2018Pattle}; Rosette: \citealt{2021Konyves}; NGC 6334: \citealt{2021Arzoumanian}; Mon R2: \citealt{2022Hwang}) and the ongoing BISTRO-3 program for various evolutionary stages and environments of star formation.
In addition to individual target studies, the BISTRO data have been used to study the polarization properties of dust grains \citep{2019Pattle,2022Fanciullo} and the alignment between magnetic fields and outflows \citep{2021Yen}.

This paper is organized as follows: in Section 2, we describe the observations and data reduction; in Section 3, we present the results of the observations; in Section 4, we derive the magnetic field strength and study the relative orientation between magnetic field and filament structure; in Section 5, we discuss our results; and in Section 6, we provide a summary of this paper.

\begin{figure*}
\centering
\includegraphics[scale=1.85]{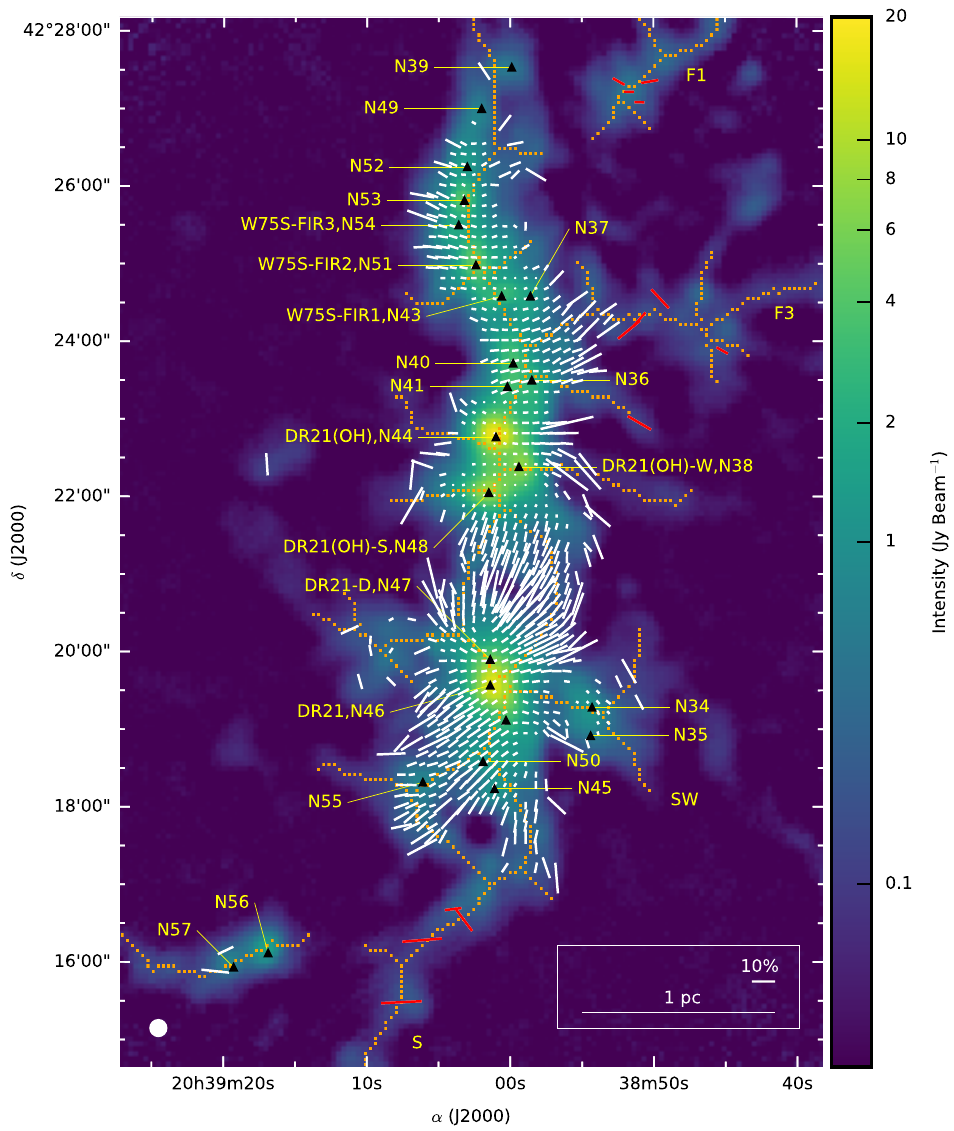}
\caption{The POL-2 dust polarization map at 850 $\mu$m toward the DR21 filament. 
The color scale represents the Stokes $I$ intensity. The magnetic field segments plotted in an interval of 8$\arcsec$ show the magnetic field orientations with the lengths proportional to the polarization percentages.
The JCMT 14.1$\arcsec$ beam is plotted at the bottom left corner.
The positions of the 24 massive dense cores in \citet{2007Motte} are marked with filled black triangles and labelled in yellow, and the filamentary structures selected using \textit{filfinder} are marked with orange dots along their crests. The names of the sub-filaments following \citet{2012Hennemann} are labeled in yellow. The magnetic field segments of the sub-filaments are shown in red color.
}
\label{fig_pol2}
\end{figure*}

\section{Observations}\label{sec_obs}
The JCMT polarization observations toward the DR21 filament were made by inserting the POL-2 polarimeter \citep{2011Bastien, 2016Friberg} into the optical path of the Submillimetre Common-User Bolometer Array 2 (SCUBA-2) camera \citep{2013Holland}. 
The observations were carried out with 20 sets of 42-minute integration in Grade 1 weather ($\tau_{\rm 225 GHz} < 0.05$) from July 2017 to February 2020 as part of the BISTRO-2 program (project ID: M17BL011). 
The observations were made using the POL-2 DAISY scan mode \citep{2016Friberg}, producing a fully sampled circular region of 12 arcmin diameter. Within the DAISY map, the noise is lowest and close to uniform in the central 3 arcmin diameter region, and increases to the edge of the map. 
The Flux Calibration Factors (FCFs) of SCUBA-2 at 850 $\mu$m were 516 Jy pW$^{-1}$ beam$^{-1}$ from November 2016 to June 2018 and 495 Jy pW$^{-1}$ beam$^{-1}$ post June 2018 \citep{2021Mairs}. Owing to the transmission losses from POL-2, the FCF of POL-2 is 1.35 times larger than the SCUBA-2 FCF \citep{2013Dempsey}. Weighted by the dates of the observations, the FCF of the POL-2 data toward the DR21 filament is 672  Jy pW$^{-1}$ beam$^{-1}$.
The effective beam size of JCMT is 14.1$\arcsec$ at 850 $\mu$m \citep{2013Dempsey}, equivalent to 0.096 pc or 2.0 $\times$ 10$^4$ AU at the distance of DR21 filament.

The data were reduced using the \textit{pol2map} procedure \citep[software version on 2020/09/22]{2018Parsons} within the STARLINK/SMURF package \citep{2013Jenness, 2014Currie}. The details of data reduction with \textit{pol2map} are described in the earlier POL-2 works such as \citet{2019Liu} and \citet{2019Wang}. In brief, the \textit{pol2map} procedure first creates an initial Stokes $I$ map from the POL-2 raw bolometer timestreams. Next, \textit{pol2map} runs a second time with fixed-signal-to-noise-based masks generated from the initial Stokes $I$ map to create improved Stokes $I$ maps and co-adds the maps into a final Stokes $I$ map.
Finally, the masks and the final Stokes $I$ map are used in a third run of \textit{pol2map} to correct instrumental polarization and produce Stokes $Q$ and $U$ maps, along with their variance maps, and the debiased polarization catalogue. The noise levels in the Stokes $Q$ and $U$ maps are estimated from the Stokes $Q$ and $U$ variance maps, which are about 3.1 mJy beam$^{-1}$ on the default 4$\arcsec$ pixels of \textit{pol2map}. The average and maximum of the noises in the Stokes $I$ map are 3.4 and 13.8 mJy beam$^{-1}$, respectively.
In this paper, we select polarization detections with criteria of $I / \delta I \geq 3$, $p/\delta p \geq 3$, and $\delta p  \leq 4\%$ for the uncertainty $\delta I$ in Stokes $I$ emission, the polarization fraction $p$, and the uncertainty $\delta p$ in $p$. We plot the polarization segments with a 90$^\circ$ rotation to show the magnetic field orientation projected on the plane of the sky (hereafter magnetic field segments), and we present one magnetic field segment in every two pixels, satisfying the Nyquist sampling of the 14.1$\arcsec$ beam.

To show the improvement of the POL-2 data, we also used the SCUPOL 
 850 $\mu$m polarization data of the DR21 filament. \citet{2009Matthews} built SCUPOL legacy catalog to provide reference Stokes cubes of comparable quality for 104 star-forming regions, including the observations of the DR21 filament of \citet{2006VF}. We downloaded SCUPOL Stokes $I$, $Q$, and $U$ cubes of DR21 from the legacy online catalogue\footnote{https://www.cadc-ccda.hia-iha.nrc-cnrc.gc.ca/en/communi-ty/scupollegacy/}. 
When comparing the POL-2 and SCUPOL data sets, we first regrided the POL-2 data to a pixel size of 10$\arcsec$ to match the SCUPOL map and then used the same criteria of $I / \delta I \geq 3$, $p/\delta p \geq 3$, and $\delta p  \leq 4\%$ to select polarization segments for both data sets, instead of the original criteria of $p / \delta p > 2$ in \citet{2009Matthews}. 
\begin{figure*}
\centering
\includegraphics[scale=1.7]{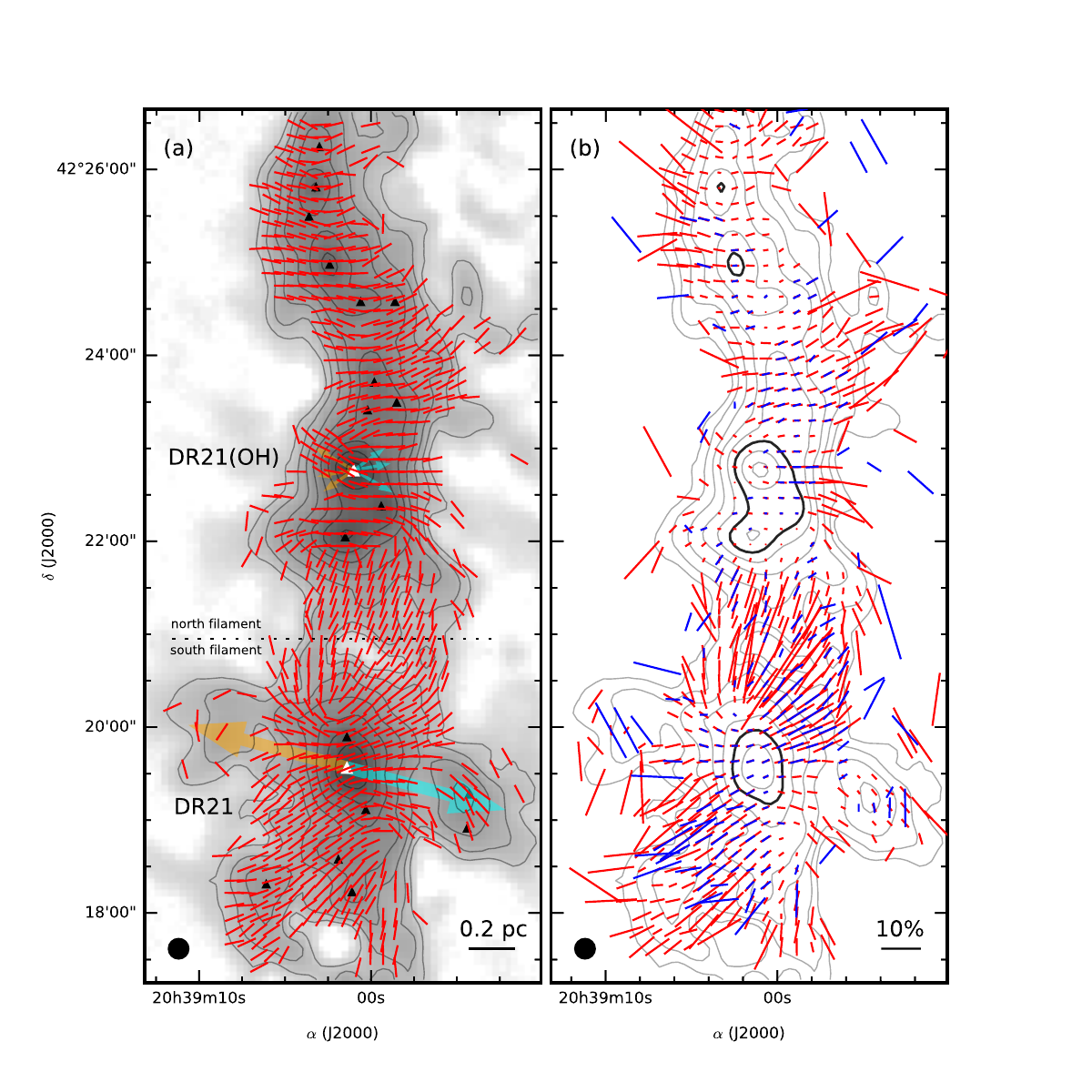}
\caption{Comparison of  the dust polarization maps between POL-2 and SCUPOL observations.
(a) The POL-2 polarization map of the main filament. The gray scale represents the Stokes $I$ intensity, and the contours show the Stokes $I$ emission at levels of 0.125, 0.25, 0.5, 1, 2, 4, 8, and 16 Jy beam$^{-1}$. The magnetic field segments are the same as those in Figure \ref{fig_pol2}, but plotted in an unified length. The triangles mark the positions of the massive dense cores in \citet{2007Motte} with the DR21(OH) in the north and the DR21 in the south highlighted in white color. 
The dotted line remarks the boundary between the north filament and south filament.
The orange and cyan arrows represent the directions of the red-shifted and blue-shifted outflows of DR21(OH) and DR21. Note here we only show the energetic outflows that might distort the POL-2 magnetic field segments in spite of the large number of outflows from the massive dense cores of the DR21 filament \citep[e.g.][]{2007Motte, 2013Zapata, 2018Ching}.
The JCMT 14.1$\arcsec$ beam is plotted at the bottom left corner.
(b) The SCUPOL magnetic field segments in blue overlapped with the POL-2 magnetic field segments in red. The length of the segment is proportional to the polarization percentage. The contours are the same as panel (a). The sixth contour at 4 Jy beam$^{-1}$ is emphasized to show the regions with high consistency between the POL-2 and SCUPOL segments.}
\label{fig_pol2scupol} 
\end{figure*}

\section{Results}\label{sec_result}

\subsection{POL-2 dust polarization map}
\subsubsection{Magnetic field morphology}
Figure \ref{fig_pol2} presents the magnetic field segments of the DR21 filament inferred from our POL-2 observations. 
The detection of dust polarized emission is more extended than the results of \citet{2006VF} and \citet{2009Matthews}, owing to a better sensitivity and a larger scan area of our observations. 
The Stokes $I$ emission shows the DR21 main filament elongated in the north-south direction embedded with the bright sources DR21(OH) and DR21 in the middle and in the south of the filament. In the eastern and western sides of DR 21, the two lobes of dust emission extend to a size of about 0.5 pc, comparable to the morphology of the energetic outflows from DR21 \citep{1996DS, 2010White}. The western side of the main filament is connected by the east-west elongated F1, F3, and SW sub-filaments, and the southern end of the filament is connected by the S sub-filament in the south-east direction. 

The magnetic field segments in the north of DR21(OH) are mostly horizontal to the filament, implying a parsec-scale magnetic field perpendicular to the main filament.
The horizontal magnetic fields are significantly changed to a northwest-southeast orientation in the region between DR21(OH) and DR21.
The magnetic fields appear to be radial around DR21 and become arc-like in the two lobes of outflows.
The arc-like morphologies of dust polarization are similar to those obtained from the imaging polarimetry of H$_2$ $v =$ 1--0 S(1) line, which suggests a helical structure of magnetic fields wrapping around the outflows \citep{1999Itoh}.
In the diffuse region, the magnetic fields of the sub-filaments are smoothly connected to the magnetic fields of the main filament. At the junctions of the sub-filaments and main filament, the magnetic fields appear to be parallel to the structures of the junctions.

Figure \ref{fig_pol2scupol}a shows a zoom-in of the polarization map to reveal the detailed magnetic field structures of the main filament.
In the north of DR21(OH), the horizontal magnetic fields are inclined in a northeast-southwest orientation in the eastern side of the filament and inclined in a northwest-southeast orientation in the western side. 
The inclined field morphology in the eastern and western sides of the filament is probably driven by the mass accretion of the filament.
In addition, the orientation and morphology of the inclined magnetic fields in the northwest of the main filament appear to be correlated with those of the F1 and F3 sub-filaments. 
The magnetic fields around massive dense cores primarily follow the horizontal magnetic fields of the filament, except for the northeast-southwest oriented magnetic fields around DR21(OH).
The northeast-southwest orientation of the magnetic fields around DR21(OH) are consistent with the small-scale magnetic fields inferred from interferometric observations of dust polarization \citep{2003Lai, 2013Girart}, and we speculate that the distortion of the magnetic fields around DR21(OH) could be driven by the northeast-southwest bipolar outflows of DR21(OH) \citep{2010White, 2012Zapata, 2013Girart}. 
At the southern end of DR21(OH), the field morphology is slightly northwest-southeast oriented along the connecting bridge between DR21(OH) and DR21.
The magnetic field morphology along the connecting bridge is probably regulated by the competitive mass accretion between the two massive cores. Because DR21(OH) is less massive than DR21, the magnetic fields in the southern end of DR21(OH) are pulled toward DR21, generating the fields that are straightened and redirected toward DR21 in a northwest-southeast orientation. 
The magnetic fields between DR21(OH) and DR21 regulated by competitive mass accretion appear to be similar to the field morphology between the massive cores in the W51 region \citep{2018Koch}.
Around DR21, the magnetic fields show a pinched or hourglass morphology with an axis of symmetry along the northwest-southeast direction, consistent with the magnetic field structure inferred from the 350 $\mu$m dust polarization observations \citep{2009Kirby, 2010Dotson}.

\begin{figure}
\centering
\includegraphics[scale=0.45]{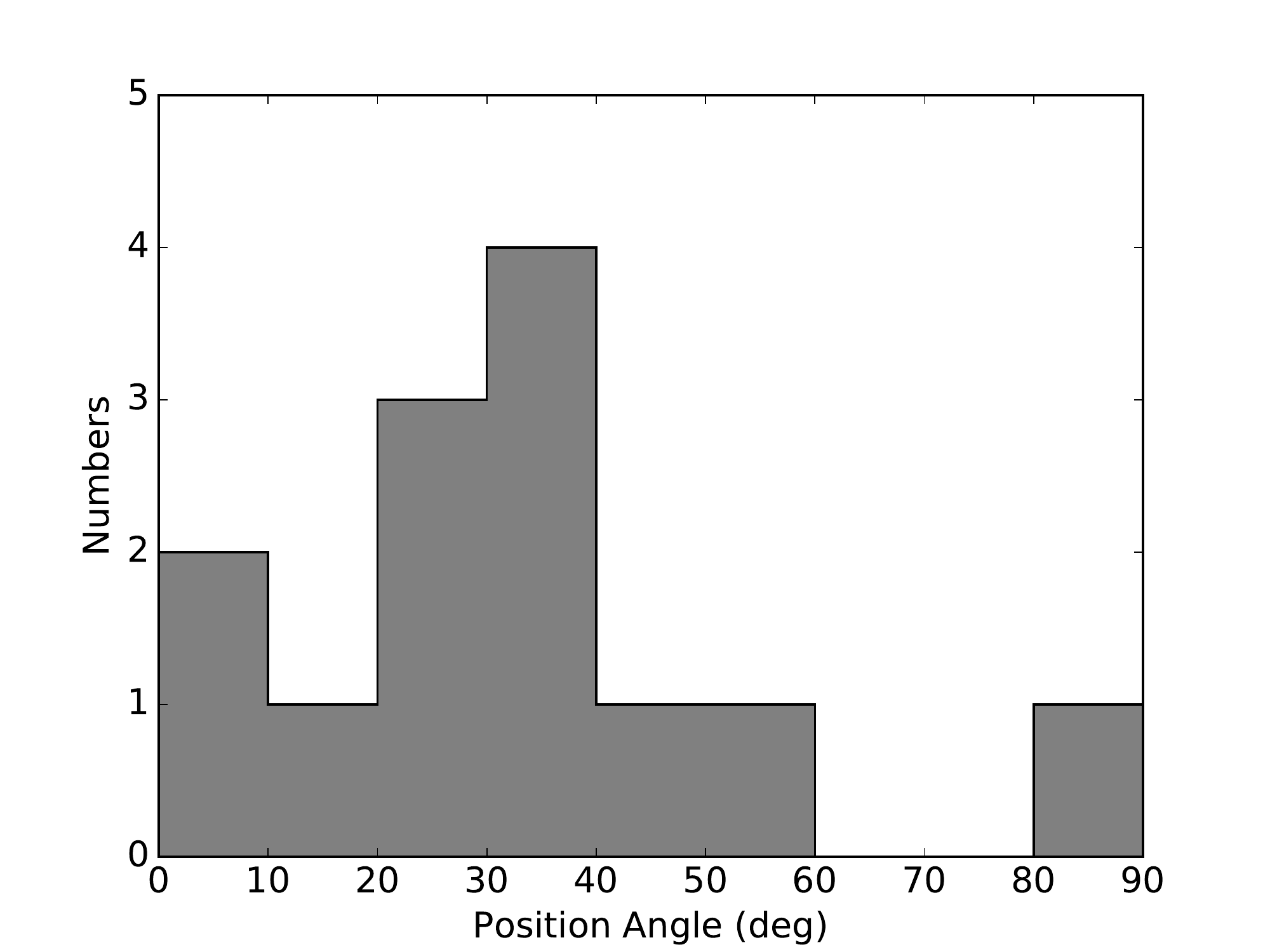}
\caption{Histogram of position angles ($PA_{|B -f|}$) between the magnetic field segments of sub-filaments and the crests of sub-filaments in Figure \ref{fig_pol2}. A position angle of 0$^\circ$ means that the magnetic field is parallel to the sub-filament crest, and a position angle of 90$^\circ$ means that the magnetic field is perpendicular to the sub-filament crest.}
\label{fig_subpolhist} 
\end{figure}

There are 13 magnetic field segments located in the sub-filaments, shown in red segments in Figure \ref{fig_pol2}. We performed  the \textit{filfinder} algorithm \citep{2015KR} to identity the crests of sub-filaments with parameters of a global threshold of 30 mJy beam$^{-1}$, a size threshold of 100 square pixels to extract filaments with length down to 40$\arcsec$ (0.3 pc), a branch threshold of 7 pixels to minimize the length for a sub-filament to be 2 beams.
The crests identified by \textit{filfinder} are plotted in Figure \ref{fig_pol2}, and the identifications of sub-filaments F1, F3, SW, and S are consistent with those in \citet{2007Kumar} and \citet{2012Hennemann}.
In Table \ref{ms_table}, we list the positions, the position angles of magnetic fields ($PA_B$), the position angles of sub-filaments ($PA_f$), and the absolute position angles ($PA_{|B -f|}$) between $PA_B$ and $PA_f$ of the 13 magnetic field segments of sub-filaments.
The $PA_f$ is determined by the five pixels of crests that are closest to the magnetic field segment.
Figure \ref{fig_subpolhist} shows the histogram of $PA_{|B -f|}$.
The histogram of the position angles has more samples between 0$^\circ$ and 45$^\circ$ than between 45$^\circ$ and 90$^\circ$, indicating that the magnetic fields tend to be parallel to the crests of sub-filaments, different to the perpendicular alignment between the magnetic fields and the DR21 main filament. 
The parallel alignment between magnetic fields and sub-filaments revealed in our POL-2 data is in agreement with the comparison of Herschel and Planck data that trace the S sub-filament and magnetic fields at a larger scale \citep{2021Hu}.

\begin{deluxetable}{ccrrc}
\tabletypesize{\scriptsize}
\tablecaption{Magnetic Field Segments of Sub-filaments\label{ms_table}}
\tablewidth{0pt}
\tablehead{
\\
$\Delta \alpha^a$  &  $\Delta \delta^a$ & $PA_B$ & $PA_f$ & $PA_{|B -f|}^b$ \\
($\arcsec$) & ($\arcsec$) & (deg) & (deg) & (deg) }
\startdata
-92 & 360 & 60.5 & 59.0 & 1.5 \\
-116 & 360 & -80.3 & -51.3 & 29.0 \\
-100 & 352 & 88.5 & -45.0 & 46.5 \\
-108 & 344 & 89.5 & -38.7 & 39.2 \\
-124 & 192 & 42.8 & 38.7 & 4.1 \\
-108 & 176 & -41.5 & -68.2 & 26.7 \\
-100 & 168 & -50.6 & -68.2 & 17.6 \\
-172 & 152 & 60.8 & 21.8 & 39.0 \\
-108 & 96 & 59.7 & 38.7 & 21.0 \\
36 & -280 & -83.1 & -51.3 & 31.8 \\
28 & -288 & 36.7 & -51.3 & 88.0 \\
60 & -304 & -85.3 & -51.3 & 34.0 \\
76 & -352 & -87.2 & -31.0 & 56.2 
\enddata
\tablenotetext{a}{With respect to the pointing center at $(\alpha,\delta)_{J2000}$ = $(20^h39^m1.1^s, +42^\circ21\arcmin17\arcsec)$}
\tablenotetext{b}{The absolute position angle between $PA_B$ and $PA_f$ in a range $\left[0^{\circ}, 90^{\circ} \right]$}
\end{deluxetable}

\subsubsection{Polarization properties}
\begin{figure}
\centering
\includegraphics[scale=0.45]{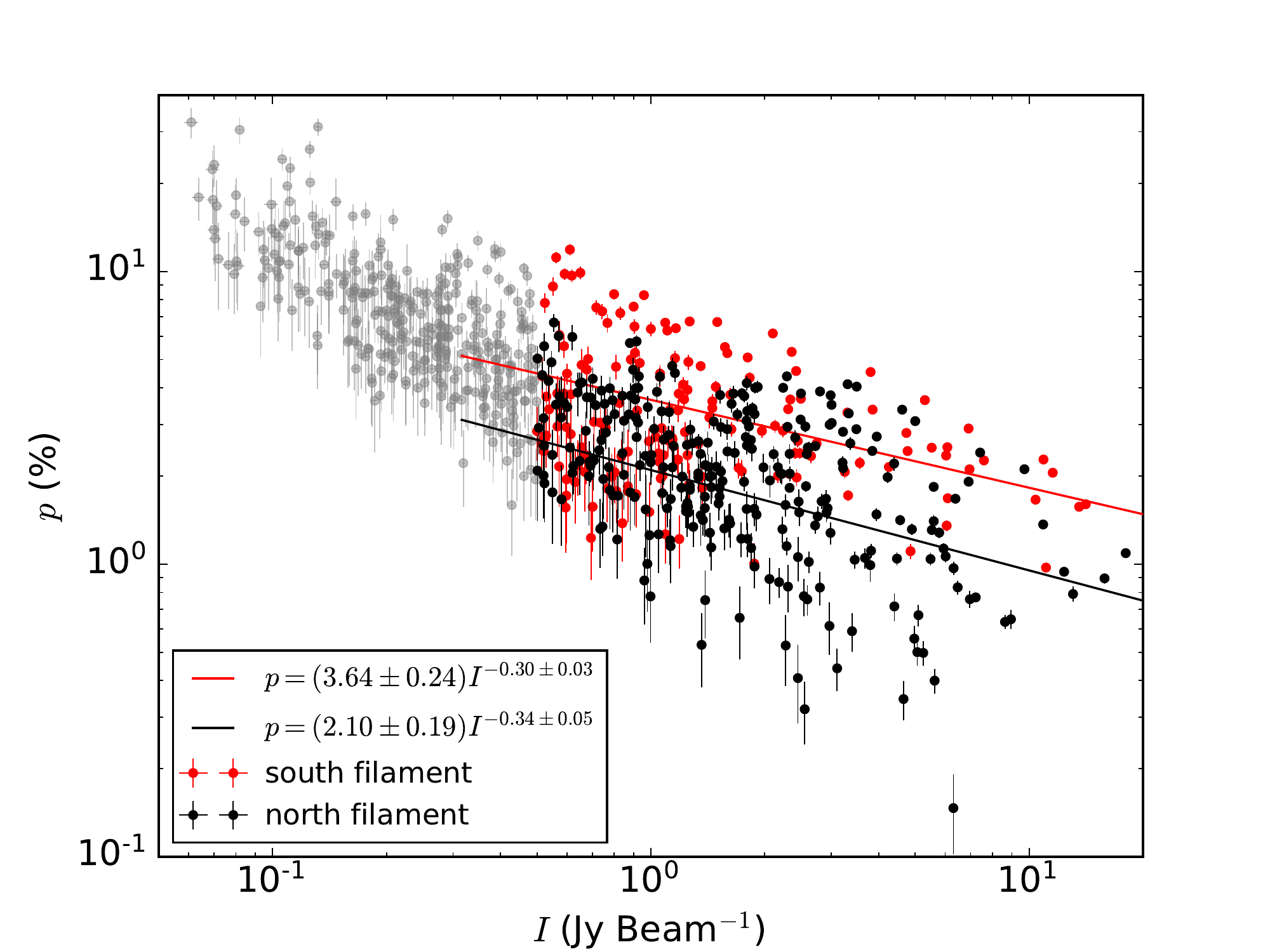}
\caption{Polarization fraction $p$ as a function of Stokes $I$ intensity.
The gray dots represent the low-intensity data ($I < 0.5$ Jy beam$^{-1}$). The black and red dots represent the high-intensity data ($I \geq 0.5$ Jy beam$^{-1}$) of the north filament and south filament, respectively. 
The best-fit models of the $p$--$I$ correlations of the north and south filaments are shown by the black and red lines, respectively.}
\label{fig_pvsi} 
\end{figure}

Figure \ref{fig_pvsi} compares the polarization fraction $p$ with the Stokes $I$ intensity for each of the POL-2 segments in Figure \ref{fig_pol2}. 
There is an overall decreasing correlation of $p$ with increasing $I$, and the low-intensity data have a steeper slope in the $p$--$I$ correlation than the high-intensity data.
In addition, the polarization fractions of several low-intensity data exceed the observed maximum polarization fraction of $22^{+3.5}_{-1.4} \%$ of the Planck 850 $\mu$m data \citep{2020Planck} and the predicted maximum polarization fraction of $\sim 15 \%$ of the submillimeter emission from interstellar dust grains \citep{2009DF}.
The steep slope of $p$--$I$ correlation and large polarization fractions ($>$ 20\%) of low-intensity data can be found in other POL-2 observations \citep[e.g.,][]{2018Kwon,2018Soam,2019Pattle,2019Wang,2019Coude,2021Arzoumanian}.
When the missing flux in Stokes $I$ data is more severe than those in Stokes $Q$ and $U$ data, the missing flux issue can lead to a polarization fraction larger than the intrinsic value. 
The steep $p$--$I$ correlation and the large polarization fractions of low-intensity data thus indicate that the low-intensity data suffer more Stokes $I$ missing flux than the high-intensity data (see Section 3.3 for a further analysis of the total and polarized missing flux in POL-2 data).

For the high-intensity data ($I \geq 0.5$ Jy beam$^{-1}$) in Figure \ref{fig_pvsi}, the polarization fractions of the segments in the north of the filament surrounding DR21(OH) are lower than those in the south of the filament surrounding DR21 (see Figure \ref{fig_pol2scupol}a for the separation boundary around the saddle region of the main filament).
To study the $p$--$I$ correlation, we use an empirical power-law model \citep{1987Tamura} with
\begin{equation}
p(I) = p_1 \left( \frac{I}{\text{Jy beam}^{-1}}\right)^{-\alpha},
\end{equation}
where $p_1$ is the polarization fraction at 1 Jy beam$^{-1}$. 
The best-fit model of the north filament gives $\alpha = 0.34 \pm 0.05$ and  $p_1 = (2.10 \pm 0.19) \%$, and the best-fit model of the south filament gives $\alpha = 0.30 \pm 0.03$ and  $p_1 = (3.64 \pm 0.24) \%$.
The difference of 0.04 between the $\alpha$ of the north filament and the $\alpha$ of the south filament is less than the uncertainty of 0.06 in the difference, whereas the difference of 1.54 \% between the $p_1$ of the north and the $p_1$ of the south filaments is about five times larger than the uncertainty of 0.31 \% in the difference.
The consistent values of $\alpha$ indicate that the dust grains of the north and south filaments have a similar property, and the significant difference in the values of $p_1$ suggests that the Stokes $I$ missing flux of the south filament is larger than the north filament, perhaps owing to differences in the intensities or spatial scales of the diffuse emission in the north and south filaments.

The values of $\alpha$ inferred from POL-2 observations of several molecular clouds are usually from 0.5 to 0.9 (IC 5146: 0.56$^{+0.27}_{-0.34}, $\citealt{2019Wang}; Barnard 1: 0.85 $\pm$ 0.01, \citealt{2019Coude}; Ophiuchus B: 0.86 $\pm$ 0.03, \citealt{2019Pattle}; Ophiuchus C: 0.83 $\pm$ 0.03, \citealt{2019Pattle}; Auriga: 0.82 $\pm$ 0.03, \citealt{2021Ngoc}; Rosette: 0.49 $\pm$ 0.08 \citealt{2021Konyves}; Serpens: 0.634, \citealt{2022Kwon};).
The 0.30--0.34 shallow values of $\alpha$ of the DR21 filament are similar to the values of $0.34 \pm 0.02$ of the Ophiuchus A \citep{2019Pattle}, $0.36 \pm 0.04$ of the Orion B \citep{2021Lyo}, and $0.35 \pm 0.02$ of the NGC 6334 \citep{2021Arzoumanian}.
The shallow $\alpha$ can be explained by the more evolved nature of the DR21 filament. According to modern grain alignment theory \citep{2007LH,2016HL,2021Hoang}, dust grains are aligned by radiative torques, and the grain alignment toward the highest intensity is caused by the internal radiation from the massive central star. As a result, the embedded sources of DR21 and DR21(OH) may increase the alignment efficiency in the high-density regions, producing a shallower $\alpha$ than those found in clouds without embedded sources.

\subsection{Comparison between POL-2 and SCUPOL results}
\begin{figure}
\includegraphics[scale=1]{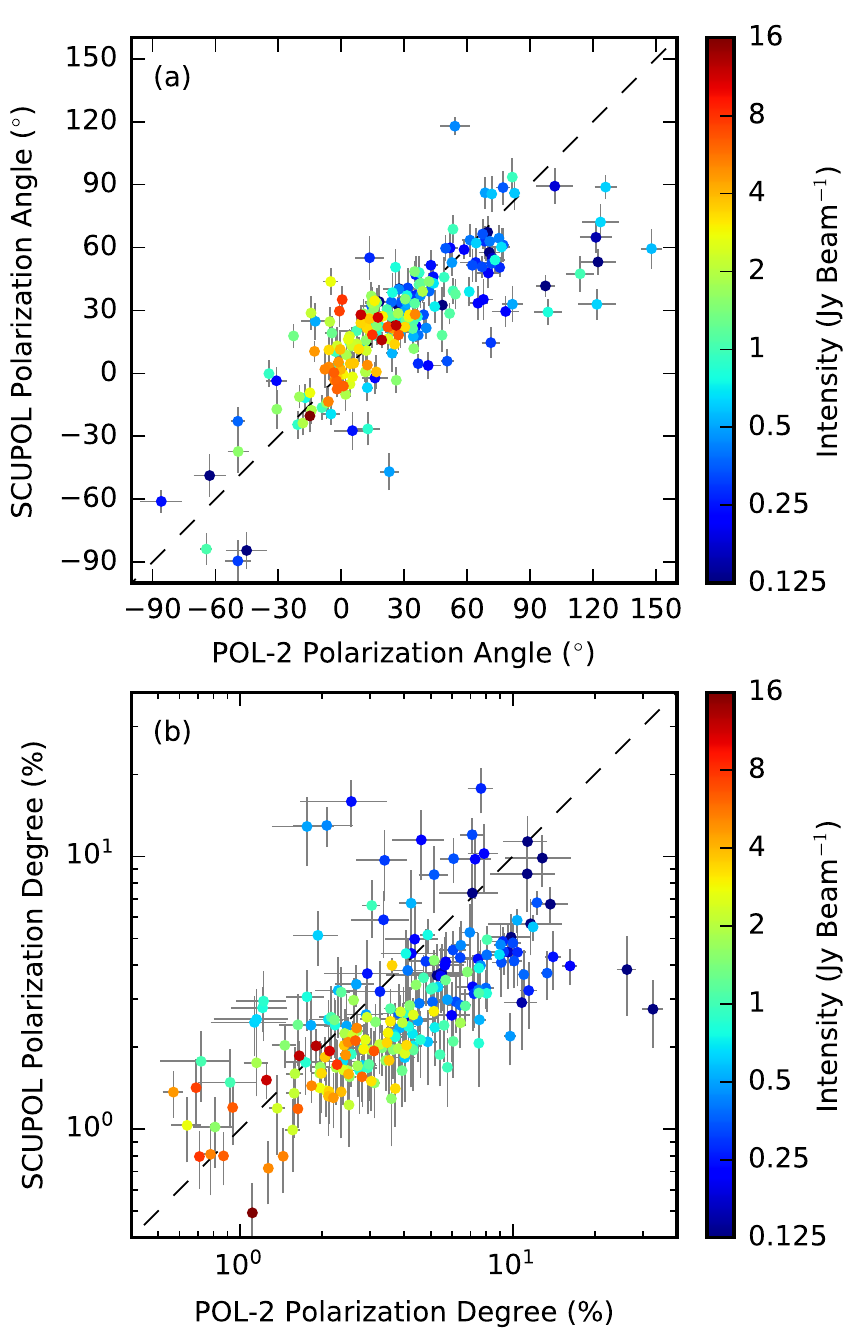}
\caption{Comparisons of polarization angles  and polarization degrees of the POL-2 and SCUPOL overlapped 215 segments in Figure \ref{fig_pol2scupol}b. The color scale represents the Stokes $I$ intensity of the data. To properly compare the polarization angles of POL-2 and SCUPOL data (i.e.\, the absolute difference between the two data sets should be less than 90$^{\circ}$) and perform a KS test,
some of the polarization angles are shifted from a range of $\left[ -90^{\circ}, 90^{\circ} \right]$ to a range of $\left[ 0^{\circ}, 180^{\circ} \right].$}
\label{fig_papmag}
\end{figure}

Figure \ref{fig_pol2scupol}b compares the polarization maps of our POL-2 data with the SCUPOL data of \citet{2009Matthews}. 
The noise level in the SCUPOL Stokes $Q$ and $U$ maps is about 13 mJy beam$^{-1}$, and that of the POL-2 data regridded to 10$\arcsec$ pixel is about 2.2 mJy beam$^{-1}$.
Above the sixth contour at 4 Jy beam$^{-1}$ intensity, the two data sets are approximately consistent in both polarization angles and polarization degrees. 
Below the sixth contour, the differences between the two data sets become larger. 
In the region between DR21(OH) and DR21 and in the south-east region of DR21, the differences in polarization angles can be as large as 50$^\circ$, and the differences in polarization degrees can be as large as 15\%.
There are 215 pairs of spatially overlapping segments between the two data sets. 
Figure \ref{fig_papmag} shows the comparisons of polarization angles and polarization degrees for the overlapping segments. 
The polarization angles and polarization degrees of the segments satisfying $I \geq$ 4 Jy beam$^{-1}$ show a better agreement between the two data sets than the segments weaker than 4 Jy beam$^{-1}$. 
The mean values of the absolute differences in polarization angles and polarization degree of the segments satisfying $I \geq$ 4 Jy beam$^{-1}$ are 8.8$^\circ$ and 0.50\%, and those values of the segments satisfying $I <$ 4 Jy beam$^{-1}$ are 18.4$^\circ$ and 2.7\%.

We further performed the two-sample Kolmogorov--Smirnov (KS) test to compare the likelihood of the POL-2 and SCUPOL polarization angles in Figure \ref{fig_papmag}a. 
When using all the data points, the KS statistic is 10.2\%, and the probability that the two samples have the same distribution at a KS significance level 0.05 is 19.9\%. 
When using the data points with $I \geq$ 4 Jy beam$^{-1}$, the KS statistic is 29.2\%, and the probability rises to 21.6\%, indicating that the POL-2 and SCUPOL sets are likely originated from the same distribution only for the high-intensity data points. 
The probability of the DR21 filament data is higher than the probabilities of 6\% in the Ophiuchus C cloud \citep{2019Liu} and 0.6\% in the Barnard 1 cloud \citep{2019Coude}.
The best consistency between the POL-2 and SCUPOL data has been found so far in the Ophiuchus B cloud with a probability of 90.5\% \citep{2018Soam}.
The KS test indicates that the consistency between the POL-2 and SCUPOL maps is better for the magnetic field segments with stronger $I$ intensities, and the improvement of POL-2 from SCUPOL in the DR21 filament is similar to the improvement of POL-2 data in other clouds.  
Considering that the sensitivity of our POL-2 data is about 3 times better than the SCUPOL data, the POL-2 segments are more reliable than the SCUPOL segments.

The $p$--$I$ correlation using the SCUPOL 439 polarization segments of DR21 filament gives $\alpha = 0.50 \pm 0.01$  \citep{2013Poidevin}. Considering that the distribution of the SCUPOL polarization segments is more extended than the lowest contour at 0.125 Jy beam$^{-1}$ in Figure \ref{fig_pol2scupol}b, the $\alpha$ derived from the SCUPOL data might be biased by the missing flux issue in the low-intensity data and therefore is steeper than our values of $\alpha$ = 0.30--0.34.

\subsection{Global magnetic fields inferred from the Planck data}
Planck 850 $\mu$m (353 GHz) polarization data are used to study the large-scale magnetic fields of the DR21 filament at the 5$\arcmin$ ($\sim$ 2.0 pc) resolution of the Planck beam. The 2015 release of Planck HFI maps \citep[PR2,][]{2016PlanckI}, where the monopole of the cosmic infrared background has been subtracted  \citep{2016PlanckVIII}, were obtained from the Planck Legacy Archive\footnote{http://pla.esac.esa.int/pla/\#home}. 
To compare the Planck and JCMT results, we transform the polarization angles of Planck data that are originally obtained in galactic coordinates into the polarization angles in equatorial coordinates by computing the angle $\psi$ between the equatorial north and the galactic north. For epoch J2000,
\begin{equation}
\psi = \arctan \left[ \frac{\cos(l-32.9^\circ)}{\cos b \cot 62.9^\circ - \sin b \sin(l-32.9^\circ)} \right],
\end{equation}
where $l$ and $b$ are the galactic coordinates of the object \citep[][see Appendix \ref{app_a} for the derivation]{1998Corradi}.

\begin{figure}
\includegraphics[scale=1]{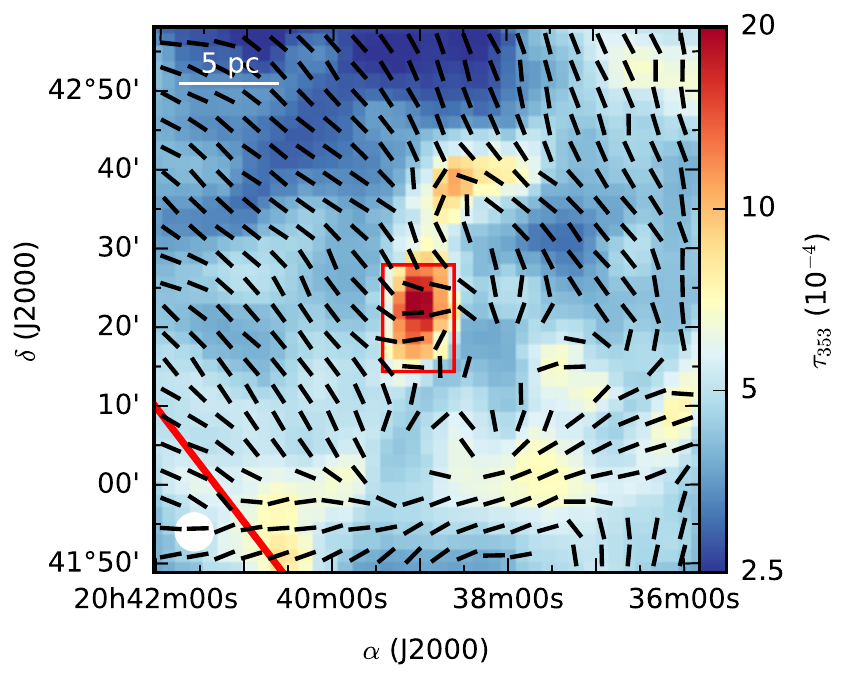}
\caption{Planck 850 $\mu$m magnetic field segments overlaid on the $\tau_{353}$ map toward the DR21 filament. The black segments represent the magnetic field orientations in an unified length. 
The red box at center remarks the area of Figure \ref{fig_pol2}.
At the bottom left corner, the red stripe represents the galactic disk plane, and the Planck 5$\arcmin$ beam is shown.} 
\label{fig_planckB}
\end{figure}

Figure \ref{fig_planckB} shows the large-scale magnetic fields inferred from Planck polarization data satisfying $p/\delta p \geq 3$ overlaid on the map of dust optical depth at 353 GHz ($\tau_{353}$) in \citet{2014PlanckXI}. 
The DR21 filament is the most prominent object in this 30 pc $\times$ 30 pc map even though DR21 is close to the galactic disk plane at $b$ $\sim$ 0.6$^\circ$.
The diffuse regions on the east side, north side, and in the northwest corner of the map show a fairly regular global magnetic field with a northeast-southwest orientation, parallel to the galactic disk plane.
This regular field is distorted in the medium above an intermediate optical depth of $\tau_{353} \sim 7 \times 10^{-4}$ in the south of the map, probably owing to the active star-forming activity of the Cygnus X complex. 
Toward the DR21 filament, a bent morphology of magnetic fields is notable: the northeast-southwest oriented global field is bent to an east-west orientation in the middle of the filament and bent to a northwest-southeast orientation in the south end of the filament, consistent with the main features of the magnetic fields of the DR21 filament at the 0.1 pc resolution of Figure \ref{fig_pol2}. 
The polarized flux in the 1$\farcm$7-sized central pixel of Figure \ref{fig_planckB} is 3.81 mK$_{\rm CMB} \times 1$\farcm$7^2= 253$ mJy, and the integrated POL-2 polarized flux over the identical area after smoothing Figure \ref{fig_pol2} to a resolution of 5$\arcmin$ is 240 mJy.
The consistency of the Planck polarized flux and POL-2 polarized flux indicates that the east-west oriented magnetic field at the center of Figure \ref{fig_planckB} is primarily traced by the polarized emission of the filament rather than by the polarized emission of the diffuse region.
Therefore, the bent morphology of large-scale magnetic fields toward the DR21 filament is associated with the 0.1-pc-scale magnetic fields of the filament rather than a distortion of large-scale magnetic fields in the diffuse region.
 
The Stokes $I$ flux in the central pixel of Figure \ref{fig_planckB} is 0.525 K$_{\rm CMB} \times 1$\farcm$7^2= 34.9$ Jy, whereas the integrated POL-2 Stokes $I$ flux over the identical area is 18.2 Jy. 
The missing large-scale flux of the POL-2 Stokes $I$ data is more severe than the Stokes $Q$ and $U$ data.
A similar trend of more missing flux in Stokes $I$ than $Q$ and $U$ is found in the POL-2 and Planck data of NGC 1333 \citep{2020Doi}.
The POL-2 Stokes $I$ missing flux of NGC 1333 is 13\% of the Planck flux, and the missing flux of DR21 filament is 48\%.
The missing flux of POL-2 data comes from the background subtraction of atmospheric signal in the \textit{pol2map} procedure, making POL-2 data not sensitive to diffuse emission with spatial scales larger than the size of the observed region.
Since the diffuse emission of DR21 filament is stronger than that of NGC 1333, the missing flux of DR21 filament hence is larger than the missing flux of NGC 1333.
In Figure \ref{fig_papmag}b, the polarization degrees of the POL-2 data are preferentially larger than those of the SCUPOL data, contrary to the general results of slightly smaller polarization degrees of POL-2 data than those of SCUPOL data \citep{2018Soam,2020Doi}. 
Again, the large POL-2 polarization degrees of DR21 filament are likely caused by the large missing Stokes $I$ flux in the POL-2 data.

\section{Analysis}\label{sec_analysis}
\subsection{Angular Dispersion Function}
\subsubsection{Formalism}
To estimate the magnetic field strength in molecular clouds from dust polarization observations,  the Davis--Chandrasekhar--Fermi \citep[hereafter DCF,][]{1951Davis, 1953CF} equation is the most widely used method.
The DCF equation assumes that the ratio of turbulence to magnetic field strength would lead to a similar level of variation in the magnetic fields as well as in the velocities, ${\delta B}/{B} \simeq {\delta V_{los}}/{V_A}$, where $B$ is the strength of the magnetic field, $\delta B$ is the variation about $B$, $\delta V_{los}$ is the velocity dispersion along the line of sight, and $V_A$ = $B$/$\sqrt{4\pi\rho}$  is the Alfv$\acute{e}$n speed at density $\rho$.
Since dust polarization segments trace the plane-of-sky component of magnetic field, the variation in the plane-of-sky magnetic field strength is expected to be proportional to the measured dispersion of polarization angles, i.e., $\delta B/B_{\rm pos} \sim \delta \Phi $.
Consequently, the DCF equation can be written as
\begin{equation}
B_{\rm pos} = F \sqrt{4\pi\rho} \frac{\delta V_{los}}{\delta \Phi},
\label{eq_dcf} 
\end{equation}  
where $F$ is a correction factor usually assumed to be $\sim$ 0.5, accounting for the smoothing of magnetic fields along the line of sight and the inadequate spatial resolution of dust polarization observations \citep{2001Heitsch, 2001Ostriker, 2001Padoan}.

To avoid inaccurate estimation of $\delta B/B_{\rm pos}$ from simply taking the dispersion of polarization angles, refinements of the DCF equation with more sophisticated statistical analyses have been made. \citet{2009Hildebrand} proposed a structure function analysis of the polarization angle difference between every pair of polarization segments in a given map as a function of the segment separation. In this structure function analysis, the plane-of-sky magnetic field is assumed to be composed of a large-scale ordered component $B_0$ and a small-scale turbulent component $B_t$, and the ratio of $B_0$ to $B_t$ can be fitted without a priori assumption on the turbulence in the cloud or the morphology of the large-scale field. \citet{2009Houde} proposed an angular dispersion function method to expand the structure function analysis by including the signal integration across the telescope beam and through the line-of-sight depth of the source. 
Recently, the method of \citet{2009Houde} has become well recognized in deriving magnetic field strength from dust polarization maps of single-dish observations \citep{2019Chuss, 2019Liu, 2019Coude, 2019Wang, 2019Soam, 2020Eswaraiah, 2021Guerra} and numerical simulations \citep{2021Liu}. 

\citet{2009Houde} suggest that if the correlation length $\delta$ for $B_t$ is much smaller than the thickness of the cloud $\Delta^{\prime}$, 
the ratio of $B_t$ to $B_0$ can be evaluated from the angular dispersion function in the form
\begin{multline}
1 - \langle \cos\left[\Delta\Phi\left(l\right)\right]\rangle \simeq  \frac{1}{N_{cell}} \frac{\langle B_t^2\rangle}{\langle B_0^2\rangle} \times \left[1-e^{-l^2/2(\delta^2+2W^2)}\right] \\ 
+ \sum_{j=1}^\infty a_{2j}^{\prime}l^{2j},
\label{eq_adf}
\end{multline} 
where $\Delta\Phi\left(l\right)$ is the polarization angle difference between polarization segments separated by a distance $l$, 
$W$ is the beam width (i.e., the FWHM beam divided by $\sqrt{8\ln{2}}$), 
the summation is a Taylor expansion representing the structure in the $B_0$ that does not involve turbulence,
and $N_{cell}$ is the number of turbulent cells along the line of sight obtained by
\begin{equation}
N_{cell} = \frac{(\delta^2+2W^2)\Delta^\prime}{\sqrt{2\pi}\delta^3}.
\label{eq_ncell}
\end{equation}  
The turbulence component in the angular dispersion function is
\begin{equation}
b^2(l) =  \frac{1}{N_{cell}} \frac{\langle B_t^2\rangle}{\langle B_0^2\rangle} e^{-l^2/2(\delta^2+2W^2)}.
\end{equation}
Since $B_t$ is the source of perturbation in $B_0$, the $\langle B_t^2\rangle / \langle B_0^2\rangle$ derived from Equation \ref{eq_adf} provides a good approximation of the $\delta B/B_{\rm pos}$ in the DCF equation for evaluating the magnetic field strength on the plane of sky as
\begin{equation}
B_{\rm pos}= \sqrt{4\pi\rho} \delta V_{los} \left[\frac{\langle B_t^2\rangle}{\langle B_0^2\rangle}\right]^{-1/2}.
\label{eq_cf}
\end{equation}

\begin{figure*}
\centering
\includegraphics[scale=0.9]{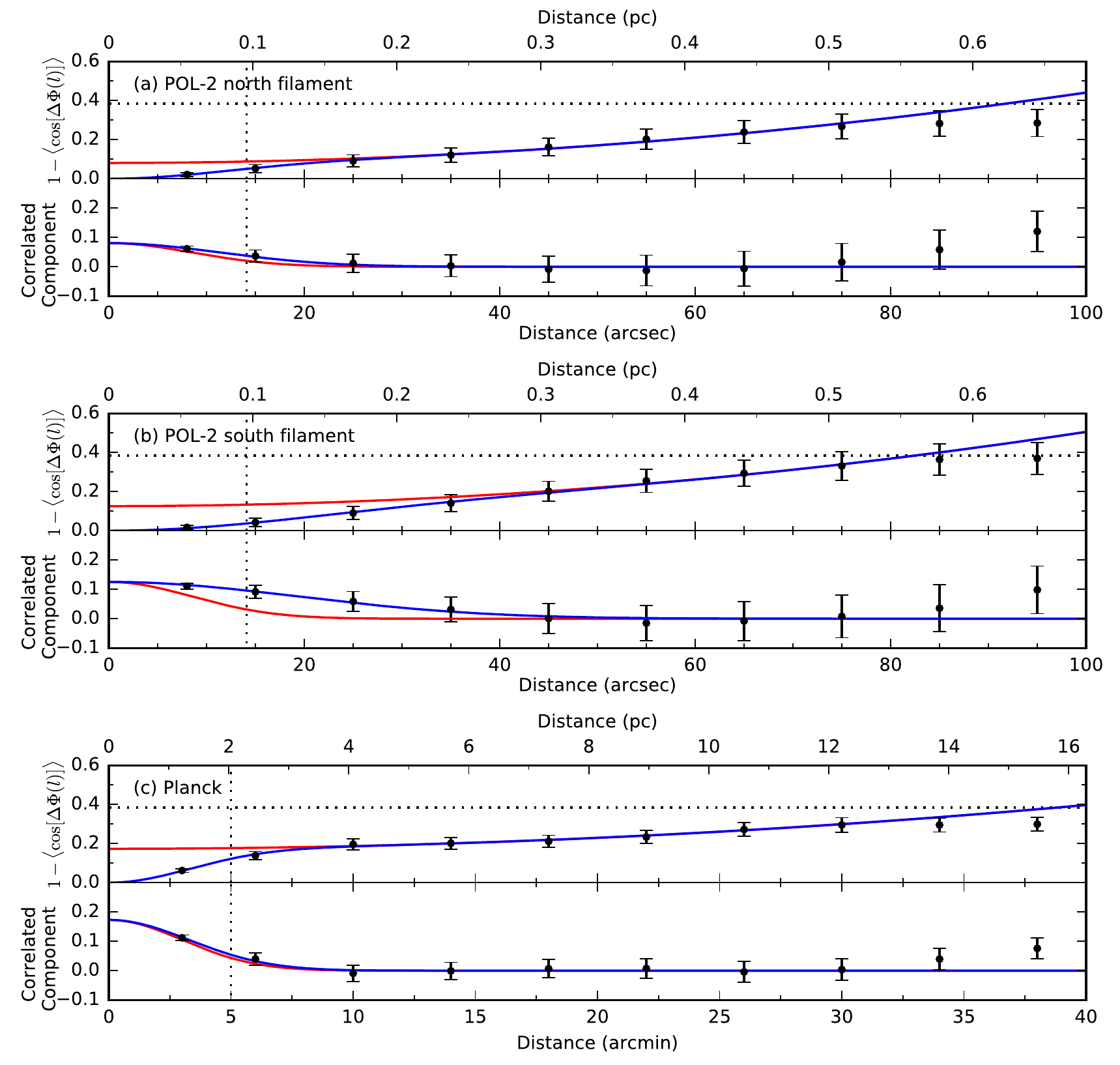}
\caption{Dispersion analysis of the POL-2 and Planck polarization segments toward the DR21 filament. For each source, the analysis of the angular dispersion function is plotted in the top panel, and the correlated component of the dispersion function is plotted in the bottom panel. 
Top panels: the dots represent the mean values of the data, and the error bars show the standard deviations of the mean values. The blue line shows the best fit to the data (Equation \ref{eq_adf}), and the red line shows the ordered component $a_2^{\prime}l^2 + b^2(0)$ of the best fit. The dotted vertical and horizontal lines denote the beam size and the expected value for random magnetic fields, respectively. 
Bottom panels: the dots represent the correlated component of the best fit to the data. The blue line shows the turbulent component $b^2(l)$ of the best fit, and the red line shows the correlation due to the beam (i.e., $b^2(l)$ when $\delta = 0$).} 
\label{fig_adf}
\end{figure*}

\subsubsection{Angular Dispersion Function of the JCMT and Planck data}
Figure \ref{fig_adf} shows the angular dispersion functions of the JCMT and Planck data toward the DR21 filament. 
Since the main feature of horizontal POL-2 segments in the north of the filament are notably different to the radial POL-2 segments in the south of the filament, we perform the analysis separately for the POL-2 segments in the north filament (Figure \ref{fig_adf}a) and in the south filament (Figure \ref{fig_adf}b). 
The numbers of segment pairs reach a maximum at $l = 80 \arcsec$ for the POL-2 data and at $l = 30 \arcmin$ for the Planck data, implying that the angular dispersion functions are fully sampled below 80$\arcsec$ for the JCMT map and fully sampled below 30$\arcmin$ for the Planck map. Here we focus on the fully sampled data points.
The POL-2 angular dispersion function in the north filament is slightly smaller than that in the south filament, indicating that the magnetic fields in the north are more ordered than those in the south.
Owing to limited angular resolution, the angular dispersion function is close to zero when the length scale $l$ is smaller than the beam.
At scales above the beams, the angular dispersion functions of POL-2 and Planck data are both at a level between 0.2 and 0.3, indicating that the ratio of ordered to turbulent magnetic fields remains similar from small to large scales.
In addition, all the angular dispersion functions of the POL-2 and Planck data are below the angular dispersion of a random field ($1-\cos 52^{\arcdeg} = 0.384$; \citeauthor{2010Poidevin} \citeyear{2010Poidevin}), indicating that the magnetic fields of the DR21 filament are considerably not random.
 
We use the nonlinear least-squares Marquardt--Levenberg algorithm\footnote{The \texttt{scipy.optimize} package of python} to fit the parameters of $\delta$, $\langle B_t^2 \rangle / \langle B_0^2 \rangle$, and $a_{2j}^\prime$ in Equation \ref{eq_adf}. 
The mean central width of the DR21 main filament and sub-filaments derived from the Herschel map is about 0.34 pc \citep{2012Hennemann}, and we use the width as the effective thickness $\Delta^\prime$ for both the JCMT and Planck data, assuming that the DR21 main filament is similar to an edge-on cylinder and its thickness is close to its width. 
We only fit the fully sampled data points, and the parameters $a_{2j}^\prime$ are reduced to first order $a_2^\prime$ because the fitting range is small.
The best fits of the angular dispersion functions are shown in Figure \ref{fig_adf}, and the fitted parameters are listed in Table \ref{table_adf}. 
The correlation lengths $\delta$ of the POL-2 north segments, POL-2 south segments, and Planck segments are $7\farcs5 \pm 1\farcs5, 17\farcs3 \pm 3\farcs1$, and $2\farcm5 \pm 1\farcm4$, respectively (see Table \ref{table_adf} for the $\delta$ in parsec). Except for the POL-2 north filament, the correlation lengths are not resolved by the beams. 
The $N_{cell}$ and $\langle B_t^2 \rangle / \langle B_0^2 \rangle$ of the sources are between 1.4 and 6.1 and between 0.2 and 0.5, respectively. 
Both the $N_{cell}$ and $\langle B_t^2 \rangle / \langle B_0^2 \rangle$ suggest that the magnetic fields more ordered than disturbed by turbulence, as the POL-2 segments in Figure \ref{fig_pol2scupol} are dominated by ordered magnetic fields perpendicular to the main filament and the Planck segments in Figure \ref{fig_planckB} are dominated by ordered magnetic fields parallel to the galactic disk plane.

\begin{deluxetable*}{ccccccccc}
\tabletypesize{\scriptsize}
\tablecaption{Angular Dispersion Function Fit Parameters}
\tablewidth{0pt}
\tablehead{
\\ 
\multirow{2}{*}{Data} & $\delta$ & \multirow{2}{*}{$\langle B_t^2 \rangle / \langle B_0^2 \rangle $} & $a_2^\prime$ & \multirow{2}{*}{$N_{cell}$} & $n$ & $\delta V_{los}^a$ & $B_{\rm pos}^b$ & \multirow{2}{*}{$\lambda^c$} \\
 & (pc) & & (arcsec$^{-2}$)& & (cm$^{-3}$) & (km s$^{-1}$) & (mG) & }
\startdata
POL-2 North & (51 $\pm$ 10) $\times 10^{-3}$ & 0.49 $\pm$ 0.17 & $(3.6 \pm 0.3) \times 10^{-5}$ & 6.1 $\pm$ 2.6 & 4.0 $\times$ 10$^{5}$ & 1.0 & 0.63 $\pm$ 0.18 & 3.9 \\
POL-2 South & (117 $\pm$ 21) $\times 10^{-3}$ & 0.18 $\pm$ 0.02 & $(3.8 \pm 0.7) \times 10^{-5}$ & 1.4 $\pm$ 0.4 & 4.0 $\times$ 10$^{5}$ & 1.0 & 1.04 $\pm$ 0.13 & 2.4 \\
Planck & 1.02 $\pm$ 0.16 & 0.27 $\pm$ 0.16 & $(4.0 \pm 0.6) \times 10^{-5}$ & 1.6 $\pm$ 0.9 & 2.0 $\times$ 10$^{4}$ & 0.7 & 0.13 $\pm$ 0.04 & 0.9
\enddata
\label{table_adf}
\tablenotetext{a}{Values from single-dish H$^{13}$CO$^+$ 1--0 observations \citep{2010Schneider}.} 
\tablenotetext{b}{Assume 10\% uncertainty in $n$ and $\delta V_{los}$ to estimate the uncertainty in $B_{\rm pos}$.} 
\tablenotetext{c}{Obtained with $B_{\rm pos} = \frac{\pi}{4}B$.}
\end{deluxetable*}

To derive the magnetic field strength using Equation \ref{eq_cf}, we adopt column densities of $41.6 \times 10^{22}$ cm$^{-2}$ for the main filament and $\sim 2 \times 10^{22}$ cm$^{-2}$ (note this value is consistent with the density derived from the Planck data in Figure \ref{fig_dp_planck}) for the diffuse region obtained from the DR21 Herschel map \citep{2012Hennemann} with $\Delta^\prime = 0.34$ pc to derive the number densities $n$ of the POL-2 and Planck maps, using $\rho = \mu m_H n$ where $\mu$ = 2.86 is the mean molecular weight \citep{2013Kirk, 2015Pattle} and $m_H$ is the atomic mass of hydrogen.

We estimate the $\delta V_{lsr}$ from the velocity dispersion ($\sigma$) of the H$^{13}$CO$^+$ 1--0 data at an angular resolution of 29\arcsec, since the emission of H$^{13}$CO$^+$ is well correlated with the dust emission in the DR21 filament \citep{2010Schneider}.  
The derived $B_{\rm pos}$ strengths are $\sim$ 0.6 mG in the north filament, $\sim$ 1.0 mG in the south filament, and $\sim$ 0.1 mG in the diffuse region.
Our value of the north filament is consistent with the SCUPOL results of $B_{\rm pos} = 0.78$ mG derived using the DCF method (Equation \ref{eq_dcf}) in \citet{2006VF} and $B_{\rm pos} = 0.62$ mG derived from the angular dispersion function analysis in \citet{2013Girart}. However, our value is about 6 times weaker than the 2.8--3.9 mG in \citet{2013Poidevin} using the structure function analysis of \citet{2009Hildebrand}, mainly owing to a 10 times larger $n$ assumed in \citet{2013Poidevin}.
Our $B_{\rm pos}$ for the south filament is lower by a factor of three than the values from 2.5 to 3.1 mG derived using the DCF method from 350 $\mu$m polarization data by \citet{2009Kirby}. The difference between the two works is primarily owing to that \citet{2009Kirby} using a velocity dispersion of 4.2 km s$^{-1}$ from the HCN 4--3 line toward the DR21 core, which is about four times larger than what we used.
 From Equations \ref{eq_adf} and \ref{eq_ncell}, $1 - \langle \cos\left[\Delta\Phi\left(l\right)\right]\rangle$ is proportional to $1/\Delta^\prime \times \langle B_t^2 \rangle / \langle B_0^2 \rangle$, and hence $\Delta^\prime$ and $\langle B_t^2 \rangle / \langle B_0^2 \rangle$ are coupled.
Considering that our assumption of $\Delta^\prime = 0.34$ pc of the diffuse region is an underestimation, $\langle B_t^2 \rangle / \langle B_0^2 \rangle$ is also underestimated. Therefore, our value of 0.13 mG in the diffuse region could be an upper limit of $B_{\rm pos}$ in the Planck data.

\subsection{Histogram of Relative Orientations}
\subsubsection{Formalism}

Dust polarization orientations in molecular clouds often show correlations with the intensity gradients inferred from the dust continuum  contours \citep{1990Goodman,2011Chapman,2012Koch}.
We quantify the relative orientation of the magnetic field with respect to the column density structures of the DR21 filament using the histogram of relative orientations \citep[HRO, ][]{2013Soler}. In the HRO technique, the relative orientation angle $\phi$ between the magnetic field and the tangent to the column density contour is evaluated using
\begin{equation}
\phi = \arctan\left( \frac{{\bf B}  \times \nabla N}{{\bf B}  \cdot \nabla N} \right),
\label{eq_phi}
\end{equation}
where ${\bf B}$ is the magnetic field orientation inferred from the polarization map and $\nabla N$ is the gradient of column density, used to characterize the column density structures. Although the range of arctan function is $\left[ -90^{\circ}, 90^{\circ} \right]$, we use a range $\left[0^{\circ}, 90^{\circ} \right]$ for $\phi$ without loss of generality as suggested in \citet{2017Soler}, since the relative orientation is independent of the reference and thus $\phi$ is equivalent to $- \phi$. 
The convention of $\phi$ is equivalent to the $| 90^\circ - \delta |$ in \citet{2013Koch} that $\phi = 0^{\circ}$ indicates that the magnetic field is parallel to the tangent of the column density contour (perpendicular to the column density gradient), and $\phi = 90^{\circ}$ indicates that the magnetic field is perpendicular to the tangent of the column density contour (parallel to the column density gradient). 

To obtain an HRO, the gradients of a column density map and the magnetic field segments of a polarization map are first compared pixel by pixel to produce a map of $\phi$. Next, the map of $\phi$ is divided into bins of column densities containing an equal number of segments, and an HRO is generated for each bin to examine the change in $\phi$ with increasing column densities. For maps with small uncertainties in column densities and polarization angles, the typical propagated error in $\phi$ is usually less than 10$^\circ$. Hence, by presenting an HRO with angle bins of a width larger than the error in $\phi$, the uncertainty in the HRO is dominated by the histogram binning process. The variance in the $k$th histogram bin is given by
\begin{equation}
\sigma_k^2 = h_k \left(1- \frac{h_k}{h_{tot}}\right),
\label{eq_sigmak}
\end{equation}
where $h_k$ is the number of samples in the $k$th bin and $h_{tot}$ is the total number of samples \citep{2016PlanckXXXV}.

To evaluate the preferential relative orientation in each column density bin, the shape of the HRO is quantified using a histogram shape parameter $\xi$, defined as, 
\begin{equation}
\xi =  \frac{A_0 - A_{90}}{A_0 + A_{90}},
\label{eq_xi}
\end{equation}
where $A_0$ is the area under the histogram in the range $0^{\circ} < \phi < 22.5^{\circ}$ and $A_{90}$ is the area under the histogram in the range $67.5^{\circ} < \phi < 90^{\circ}$ \citep{2017Soler}.  An HRO peaking at $\phi = 0^{\circ}$ would have $\xi > 0$, an HRO peaking at $\phi = 90^{\circ}$ would have $\xi < 0$, and a flat HRO would have $\xi \sim 0$. The uncertainty in $\xi$ is obtained from
\begin{equation}
\sigma_\xi^2 = \frac{4(A_{90}^2 \sigma_{A_{0}}^2 + A_{0}^2 \sigma_{A_{90}}^2)}{(A_0 + A_{90})^4},
\label{eq_xi2}
\end{equation}
where $\sigma_{A_{0}}^2$ and $\sigma_{A_{90}}^2$ represent the variances of the areas, characterizing the ``jitter'' of the histograms. The value of $\xi$ is nearly independent of the number of angle bins selected to represent the histogram if the bin widths are smaller than the integration range, but the jitter does depend on the number of angle bins in the histogram. If the jitter is large, $\sigma_\xi$ is large compared to $| \xi |$, and the relative orientation is indeterminate \citep{2016PlanckXXXV}. 

Finally, analyses of HROs characterize the trend of the relative orientation between magnetic fields and column density structures of a cloud from its low to high density regions with a linear regression between $\xi$ and atomic gas column density $N(H)$ \citep{2016PlanckXXXV}: 
\begin{equation}
\xi = C_{HRO} \left[\log_{10}(N(H)/ \text{cm}^{-2}) - X_{HRO}\right].
\label{eq_CX}
\end{equation}

\begin{figure*}
\centering
\includegraphics[scale=1.7]{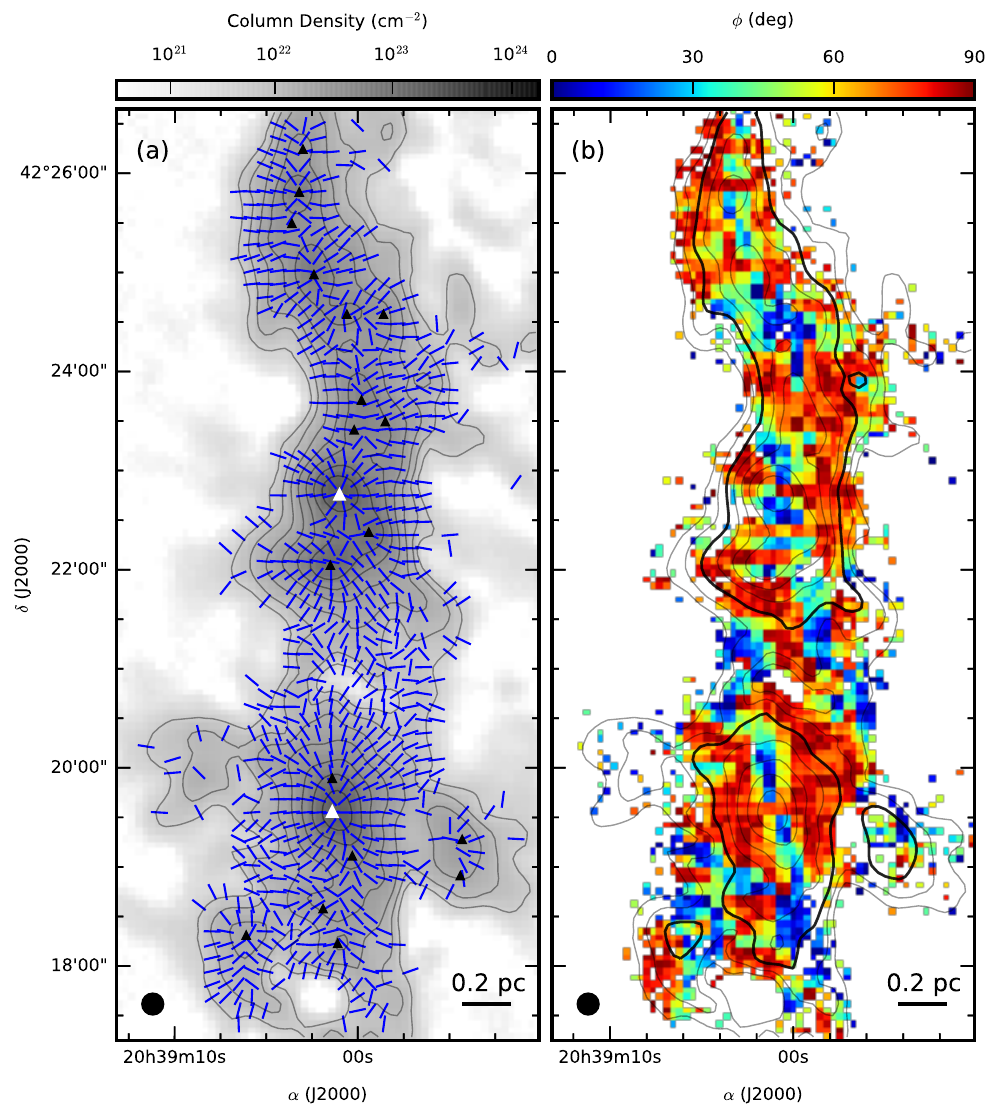}
\caption{Comparison of the magnetic field segments and column density gradient segments of the POL-2 data. (a) $N(H_2)$ column density map derived from the JCMT Stokes $I$ emission overlaid with the gradient segments calculated by convolving the column density map with a Gaussian derivative kernel. 
The contours show the $N(H_2)$ at levels of 0.125, 0.25, 0.5, 1, 2, 4, 8, and 16 $\times 10^{23}$ cm$^{-2}$.
The length of the gradient segments is normalized. The gradient segments shown here are those overlaid with the magnetic field segments in Figure \ref{fig_pol2scupol}a.
(b) The map of relative orientation angle $\phi$ between the magnetic field segments in Figure \ref{fig_pol2scupol}a and the gradient segments in panel (a).
The contours are the same as panel (a) with the contour at $5 \times 10^{22}$ cm$^{-2}$ emphasized to show the transition from no preferential orientation of $\phi$ in low density regions to perpendicular orientation of $\phi$ in high density regions.} 
\label{fig_dp_jcmt}
\end{figure*}

\subsubsection{Histogram of Relative Orientations of the JCMT and Planck data}
In order to further compare the analyses of HROs of Planck and JCMT data from low-density to high-density regimes, we construct the column density maps of the data.
To convert the JCMT dust continuum map to a column density map, we calculate the column density $N(H_2)$ of molecular gas as follows:
\begin{equation}
N(H_2) = \frac{\gamma I_\nu}{\mu m_H \kappa_\nu B_\nu (T)},
\end{equation}
where $\gamma$ is the gas-to-dust ratio of 100, $I_\nu$ is the Stokes $I$ intensity at frequency $\nu$, 
$\kappa_\nu$ = 1.5 cm$^2$ g$^{-1}$ is the dust opacity at 850 $\mu$m of cool and dense dust mantles \citep{1994OH}, and $B_\nu (T)$ is the Planck function at the dust temperature $T$ of 15 K previously measured in the DR21 filament \citep{2012Hennemann}. To scale the Planck $\tau_{353}$ map to a column density map, we calculate the column density $N(H)$ of atomic gas following the dust opacity relation found using Galactic extinction measurements of quasars \citep{2014PlanckXI},
\begin{equation}
\tau_{353}/N(H) = 1.2 \times 10^{-26} \text{cm}^2.
\end{equation}
We next calculate the gradients of the $N(H_2)$ and $N(H)$ maps using the Gaussian Derivatives method described in \citet{2013Soler}. 
To obtain gradients at the pixels of the POL-2 and Planck magnetic field segments in Figures \ref{fig_pol2} and \ref{fig_planckB}, we apply a 3 $\times$ 3 derivative kernel over the grid of pixels illustrating magnetic field segments.
Since the gradients in this grid are computed over two FWHM beams for both the JCMT and Planck data, obtaining gradients using this method guarantees adequate sampling of gradients.

\begin{figure}
\centering
\includegraphics[scale=0.9]{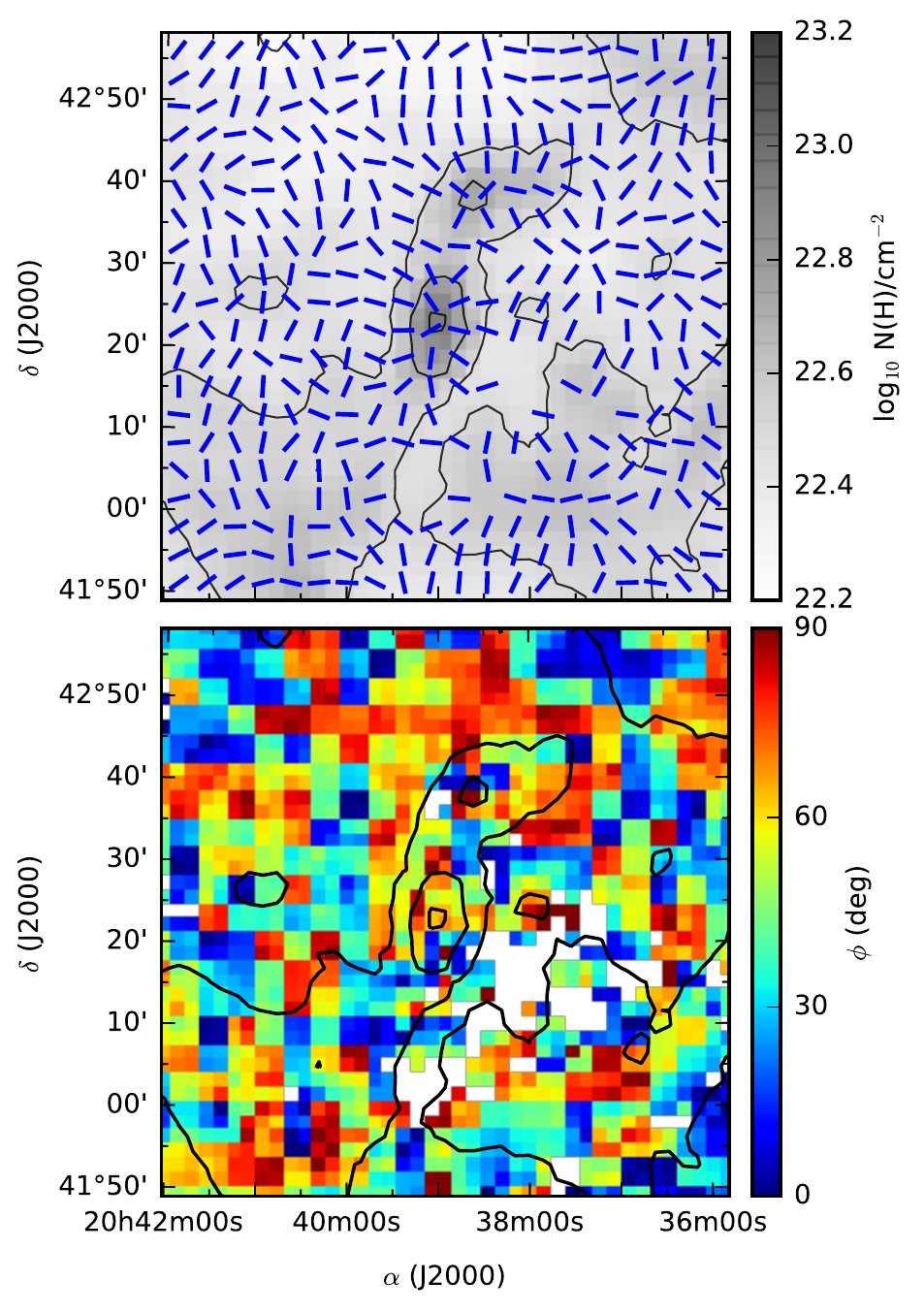}
\caption{Comparison of the magnetic field segments and column density gradient segments of the Planck data. Top panel: $N(H_2)$ column density map derived from the Planck $\tau_{353}$ map overlaid with the gradient segments calculated by convolving the column density map with a Gaussian derivative kernel. 
The contours show the $N(H_2)$ at levels of 2, 4, 8, and 16 $\times 10^{22}$ cm$^{-2}$.
The length of the gradient segments is normalized. The gradient segments shown here are those overlaid with the magnetic field segments in Figure \ref{fig_planckB}. Bottom panel: The map of relative orientation angle $\phi$ between the magnetic field segments in Figure \ref{fig_planckB} and the gradient segments in top panel.}
\label{fig_dp_planck}
\end{figure}

Figures \ref{fig_dp_jcmt} and \ref{fig_dp_planck} show the gradient segments of the column density maps and the maps of $\phi$ of the JCMT and Planck data.
The majority of POL-2 segments in the DR21 main filament tends to be parallel ($\phi = 90^{\circ}$) to the gradient segments with $N(H_2) \gtrsim 10^{23}$ cm$^{-2}$. In the low column density regions of the JCMT map, the alignment between POL-2 segments and gradient segments becomes less significant. The large-scale magnetic field segments and gradient segments in the Planck map appear to be more randomly aligned than the small-scale segments in the JCMT map. 
The uncertainty in the position angle of the gradient is determined by the derivative of the noise in the column density map \citep{2016PlanckXXXV}.
Since the respective noise levels in the POL-2 Stokes $I$ map and the Planck $\tau_{353}$ map are much less than a few percent of the $I$ map and $\tau_{353}$ values, the uncertainties in the gradient directions are typically less than 1$^\circ$. 
We use a selection criterion of $p/ \delta p \geq 3$ for the magnetic field segments, corresponding to an uncertainty less than 10$^\circ$ in polarization angle \citep{1993NC}. Therefore, we expect that the errors in $\phi$ are less than 10$^\circ$.

\begin{figure}
\centering
\includegraphics[scale=0.4]{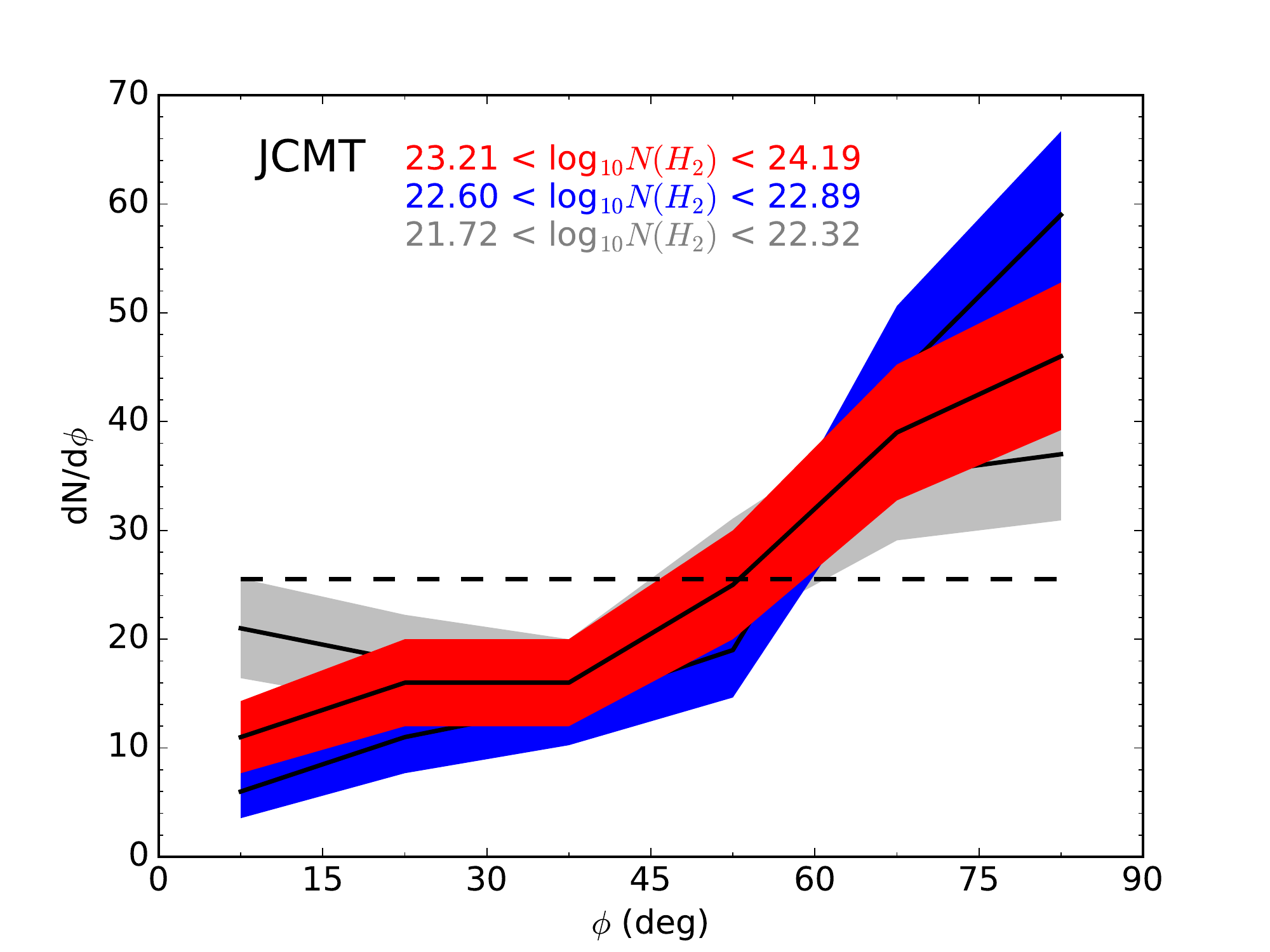}
\caption{HROs for the lowest, the intermediate, and the highest $N(H_2)$ bins (gray, blue, and red, respectively) of the JCMT data. 
The horizontal dashed line corresponds to the average HRO per angle bin of 15$^{\circ}$ for a $N(H_2)$ bin. The widths of the shaded areas for each histogram correspond to the $\pm$1 $\sigma_k$ uncertainties (Equation $\ref{eq_sigmak}$) related to the histogram binning operation.} 
\label{fig_HRO_jcmt}
\end{figure}

\begin{figure}
\centering
\includegraphics[scale=0.4]{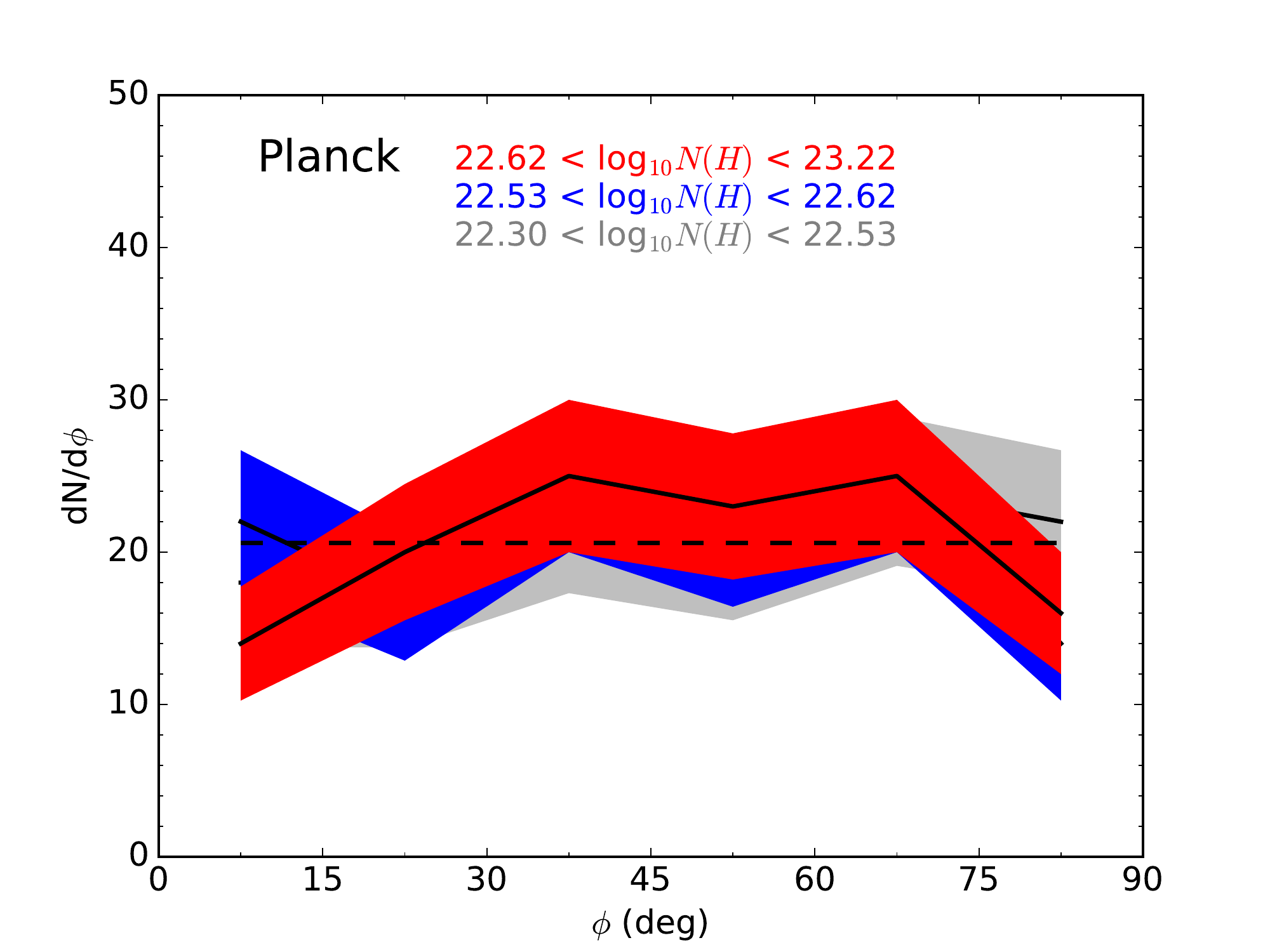}
\caption{HROs for the lowest, the intermediate, and the highest $N(H)$ bin (gray, blue, and red, respectively) of the Planck data. 
The horizontal dashed line corresponds to the average HRO per angle bin of 15$^{\circ}$ for a $N(H)$ bin. The widths of the shaded areas for each histogram correspond to the $\pm$1 $\sigma_k$ uncertainties (Equation \ref{eq_sigmak}) related to the histogram binning operation.} 
\label{fig_HRO_planck}
\end{figure}

We divide the 765 measurements of $\phi$ of the POL-2 data into 5 $N(H_2)$ bins and the 371 measurements of $\phi$ of the Planck data into 3 $N(H)$ bins to calculate HROs. Figures \ref{fig_HRO_jcmt} and \ref{fig_HRO_planck} plot the HROs of the JCMT and Planck data using 6 angle bins each of 15$^\circ$ width.
The HROs reveal different kinds of relative orientations between magnetic fields and column density contours of the JCMT and Planck data.
The JCMT HRO of the lowest $N(H_2)$ bin increases slightly from $\phi = 0^\circ$ to $\phi = 90^\circ$, and the HROs of the intermediate and highest $N(H_2)$ bins show prominent peaks at 90$^\circ$,
suggesting a trend from a weak perpendicular orientation of $\phi$ in regions with $N(H_2) \lesssim 10^{22.5}$ cm$^{-2}$ to a strong perpendicular orientation of $\phi$ for $N(H_2) \gtrsim 10^{22.5}$ cm$^{-2}$ in the DR21 main filament.
In contrast, the Planck HROs are flat for all of the three $N(H)$ bins, suggesting no preferential orientation of $\phi$ in the large-scale diffuse region of the DR21 filament.

\begin{figure}
\centering
\includegraphics[scale=0.4]{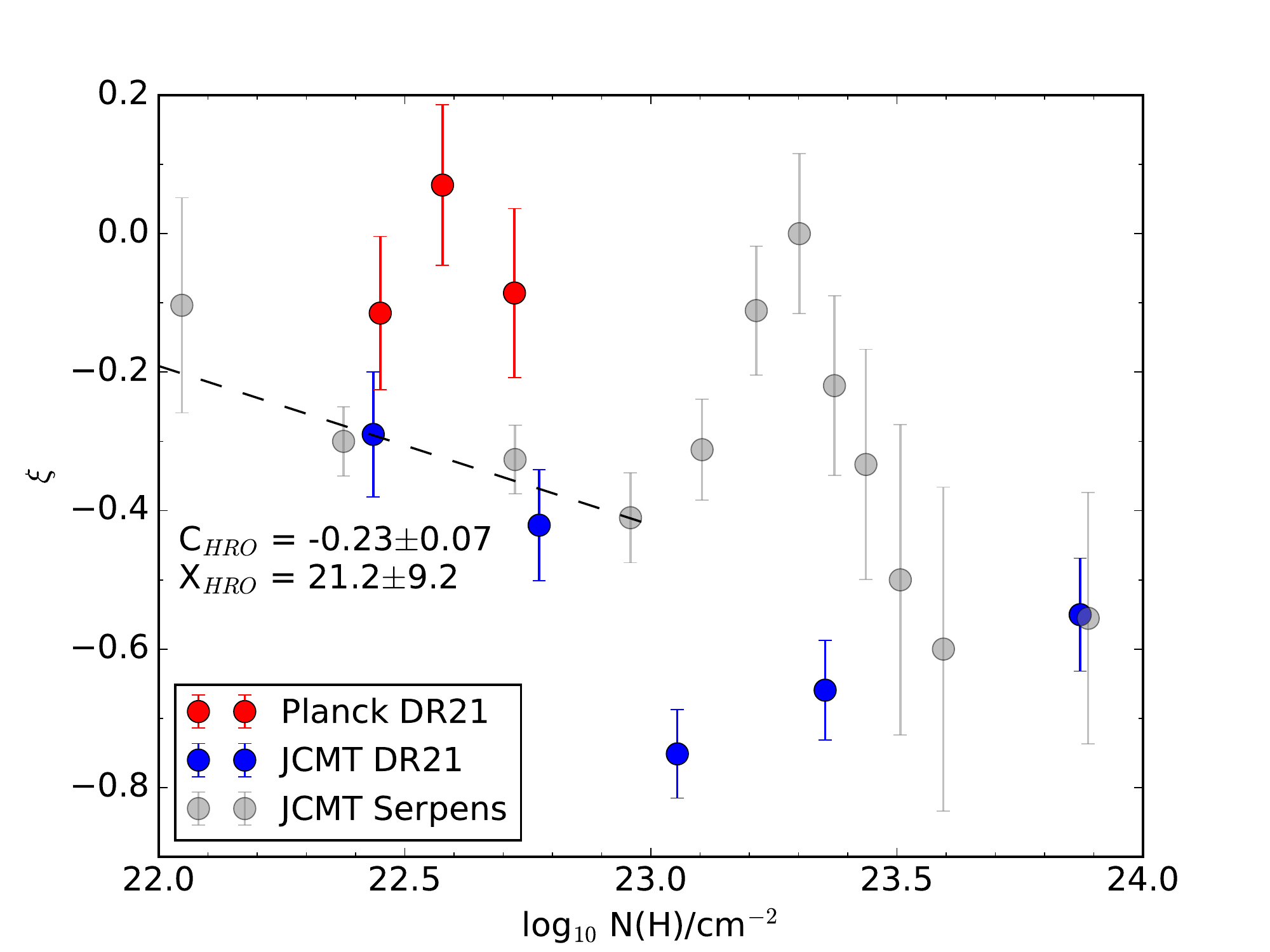}
\caption{Relative orientation parameter $\xi$, defined in Equation \ref{eq_xi} and \ref{eq_xi2}, calculated for the different $N(H)$ bins of the Planck (red) and JCMT (blue) data of the DR21 filament. The JCMT results of $\xi$ in the Serpens Main region are shown in gray dots.
The black dashed line and the values of $C_{HRO}$ and $H_{HRO}$ correspond to the linear fit of Equation \ref{eq_CX} for the JCMT data below $N(H) = 10^{23}$ cm$^{-2}$.}
\label{fig_xi}
\end{figure}

Figure \ref{fig_xi} presents the measurements of $\xi$ in different $N(H)$ bins derived from the HROs of the JCMT and Planck data. 
To compare the $\xi$ of the two data sets, the column density $N(H_2)$ is transferred to the $N(H)$ assuming $2 \times N(H_2) = N(H)$.
The $\xi$ of the three Planck $N(H)$ bins are consistently close to zero with a relatively large value of $\sigma_\xi$.
The $\xi$ of the lowest JCMT $N(H)$ bin is slightly smaller than the $\xi$ of the three Planck bins.
Considering that the $\sigma_\xi$ of the four data points are relatively large and the missing flux of the POL-2 Stokes $I$ measurement (see Section 3.3) might cause the lowest JCMT $N(H)$ bin to be smaller than its intrinsic column density, the $\xi$ of the JCMT data seems to agree with the $\xi$ of the Planck data. 
The $\xi$ for the rest of the JCMT $N(H)$ bins are broadly negative, indicating a strong preference of perpendicular alignment between the small-scale magnetic fields and the ridge of the DR21 filament (parallel alignment between magnetic fields and density gradients).

\section{Discussion}
\subsection{The Role of Magnetic Field in the DR21 Filament}
One of the important parameters required to evaluate the role of magnetic fields in star formation is the dimensionless mass-to-flux ratio $\lambda$, which refers to the ratio of the mass in a magnetic flux tube to the magnitude of magnetic flux \citep{2004Crutcher}.
In units of its critical value of $2\pi G^{1/2}$, $\lambda = 7.6 \times 10^{-21} [N(H_2)/\textrm{cm}^{-2}][B/\mu \textrm{G}]^{-1}$.
In the theory of magnetic dominated star formation \citep{1987Shu, 1999MC}, clouds are initially magnetically subcritical ($\lambda < 1$), and to become a star-forming region, magnetic supercriticality ($\lambda > 1$) of a cloud is required for the self-gravity to overwhelm the magnetic support and form stars through gravitational collapse.
Using the statistically most probable value of $B_{\rm pos} = \frac{\pi}{4} B$ \citep{2004Crutcher} and the column density obtained from the Herschel map (see Section 4.1.2), the $\lambda$ values of the main filament and diffuse region are listed in Table \ref{table_adf}.
Given the uncertainties in the column densities, in the $B_{\rm pos}$, and in the projection correction from $B_{\rm pos}$ to $B$, we estimate that the uncertainty in $\lambda$ would be as large as half of the value.
The $\lambda$ values of the main filament is about 2.4--3.9, consistent with the value of 3.4 obtained using SCUPOL data \citep{2013Girart}
and the value of 2--3 obtained using the CN Zeeman measurements toward DR21(OH) \citep{1999Crutcher}.
The $\lambda$ of the diffuse region is 0.9, however, which should be taken as a lower limit because the $B_{\rm pos}$ of the diffuse region could be overestimated.

These $\lambda$ imply different roles of magnetic fields in the main filament and the surrounding diffuse region. 
The significant supercriticality of the main filament implies that self-gravity dominates magnetic fields and the filament is undergoing gravitational collapse, in agreement with the infall motions of the filament suggested from molecular line observations \citep{2010Schneider, 2011Csengeri}.
Meanwhile, the observed perpendicular alignment between the main filament and magnetic fields is consistent with the MHD simulations of a strongly magnetized medium that magnetic field can regulate mass flows along field lines to form parsec-scale filamentary structures perpendicular to magnetic fields \citep[e.g.][]{2008NL, 2013IF, 2014CO, 2019LK}.
In contrast, the $\lambda$ of the diffuse region is slightly subcritical or nearly critical, indicating that the ambient gas is incapable to form the DR21 filament through direct gravitational collapse.
Considering that the column densities of the sub-filaments are between those of the diffuse region and main filament, the 
$\lambda$ of sub-filaments should be larger than that of the diffuse region and smaller than that of the main filament.
In other words, the sub-filaments might be the places where the transition from subcriticality to supercriticality occurs. 
The sub-filaments of DR21 appear to be parallel to the parsec-scale magnetic fields and perpendicular to the main filament.
These features are similar to the striations around filamentary clouds in MHD simulations formed via Alfv{\'e}n waves \citep{2008Heyer, 2018TT} or Kelvin--Helmholtz instability \citep{2017Chen, 2019LK}.

The magnetic fields of the DR21 filament seem to play a more important role on large scales and become less important on small scales. 
At scales of a few parsecs, the magnitude of magnetic flux is comparable to self-gravity, preventing the collapse of ambient gas. 
For the parsec-scale main filament, the magnetic fields are important in shaping the filamentary structure, even though the magnetic fields are overwhelmed by the self-gravity of the filament.
The magnetic fields of six massive dense cores, including DR21(OH), in the filament have been studied in \citet{2013Girart} and \citet{2017Ching} using dust polarization observations at resolutions of a few thousand au. 
In contrast to the ordered parsec-scale magnetic fields that are perpendicularly aligned to the filament, the magnetic fields of those cores have complex structures that appear to be randomly aligned to the core structures.
The $\lambda$ of the cores are supercritical with values comparable to that of the main filament, but the ratio of virial kinematic energy to virial magnetic energy of the cores is at least an order of magnitude larger than that of the filament. 
Meanwhile, molecular line observations suggest that increasing kinetic energy in the core comes from gravitational collapse and might be the source of the distortion of the magnetic fields into complex structures  \citep{2018Ching}.
Hence, the massive cores appear to be weakly magnetized, and self-gravity and gas dynamics are more important than magnetic fields in the formation of massive dense cores.
Down to scales of $\sim$ 1000 au, the study of fragmentation of 18 massive dense cores, including three cores in the DR21 filament, suggests that the correlation between the fragmentation levels and the number densities of the cores is stronger than the correlation between the fragmentation levels and the $\lambda$ of the cores \citep{2021Palau}.

\subsection{Comparison of the HROs of the DR21 Filament and Other Clouds}

The HRO of the Serpens Main region of the BISTRO survey has been studied in \citet{2022Kwon}. In Figure \ref{fig_xi}, the $\xi$ measurements of the Serpens Main region are overlaid on those of the DR21 filament. Both the DR21 and the Serpens Main data show a turning point of $\xi$ around $N(H)$ of 10$^{23}$ cm$^{-2}$. Owing the displacement in column density between the JCMT and Planck data, we select the $\xi$ of the JCMT data below $N(H)$ = 10$^{23}$ cm$^{-2}$ to derive the $C_{HRO}$ and $X_{HRO}$ in Equation \ref{eq_CX}.
The resulting $C_{HRO}$ of $-$0.23 reflects the trend that $\xi$ changes from a value close to zero in the low $N(H)$ bins to negative values in the high $N(H)$ bins. The resulting $X_{HRO}$ of 21.2, equivalent to $1.58 \times 10^{21}$ cm$^{-2}$, corresponds to a characteristic column density  where $\xi$ changes its sign, or in other words, a boundary where the relative orientation between the column density structures and magnetic fields changes from a more random orientation in low-density regions to a non-random, preferentially perpendicular orientation in high-density regions.

The HROs of ten nearby Gould Belt molecular clouds (at distances of less than 450 pc, namely Taurus, Ophiuchus, Lupus, Chamaeleon--Musca, Corona Australis, Aquila Rift, Perseus, IC5146, Cepheus, and Orion) have been measured using the Planck data smoothed to 10$\arcmin$ resolution \citep{2016PlanckXXXV}.
The $\xi$ of the HROs are found to decrease with increasing $N(H)$, indicating field orientation from preferentially parallel or having no preferred orientation at the lowest $N(H) \sim 10^{21}$ cm$^{-2}$ of the data to preferentially perpendicular at the highest $N(H) \sim 10^{22.5}$ cm$^{-2}$ of the data.
Except for the Corona Australis cloud that shows an almost flat slope of $\xi$, the $C_{HRO}$ of the other nine clouds have a  range from $-0.22$ to $-0.68$, and the $X_{HRO}$ have a range from $21.67$ to $22.70$.
The HROs of the ten clouds and the high-latitude cloud L1642 have further been studied between the $N(H)$ derived from Herschel data at 20$\arcsec$ resolution and the magnetic fields inferred from Planck 850 $\mu$m polarization data, and negative slopes of $\xi$ versus $N(H)$ are identified \citep{2016Malinen, 2019Soler}.
Besides the Planck polarization data, the HRO analysis applied to the BLASTPol data at 250 $\mu$m, 350 $\mu$m, and 500 $\mu$m at 3$\arcmin$ resolution toward the Vela C molecular complex with $N(H)$ from 10$^{21.7}$ cm$^{-2}$ to 10$^{23.3}$ cm$^{-2}$ also suggests a similar trend of HRO as the Planck results \citep{2017Soler}.


The $C_{HRO}$ and $X_{HRO}$ of the DR21 filament and the Serpens Main region for $N(H) < 10^{23}$ cm$^{-2}$ are $-0.23$ and 21.2, consistent with the values of other molecular clouds in the same density regime.
The observed change in the HRO from mostly parallel alignment between magnetic fields and sub-filaments of diffuse gas to mostly perpendicular alignment between magnetic fields and dense filaments of clouds is consistent with recent simulations of MHD turbulence with strong magnetic fields, indicating that magnetic fields play a significant role in structuring the interstellar medium in and around molecular clouds \citep{2013Soler, 2017SH}.
Yet, there are two features in Figure \ref{fig_xi} that are different to the HROs of most molecular clouds. 
First, because the angular resolution of POL-2 is higher than other single-dish polarimeters, the HROs of the DR21 filament and the Serpens Main region trace the highest $N(H)$ of 10$^{24}$ cm$^{-2}$.
Second, the $\xi$ of the DR21 filament and the Serpens Main region reaches a minimum between $-0.6$ and $-0.8$, which is lower than other molecular clouds, except for the HRO of Musca obtained from Herschel and Planck data in \citet{2019Soler}. 
Considering that a perfectly perpendicular alignment between magnetic field and filament would give $\xi = -1$, it seems reasonable for high angular resolution observations to obtain a $\xi$ close to $-1$ in a high-density filament, such as the DR21 filament.
  
For $N(H) > 10^{23}$ cm$^{-2}$, Figure \ref{fig_xi} shows a tentative positive slope in $\xi$ versus $N(H)$ in both the DR21 filament and the Serpens Main region. Similar trends of positive slopes in high $N(H)$ regimes can be found in the HROs of Lupus I, Musca, Perseus, and Vela C South-Nest \citep{2017Soler,2019Soler}. 
Statistical studies of magnetic fields in star-forming cores suggest that the small-scale magnetic fields of cores are neither simply aligned with the large-scale magnetic fields of filaments nor simply aligned with the major axes of filaments \citep{2014Zhang, 2014Koch}. 
Therefore, the tentative positive slope in the high $N(H)$ regime of the HRO may indicate the transition from preferentially perpendicular alignment between filaments and magnetic fields to complex structure of the alignment between dense cores and magnetic fields.
Characterizing HROs with high angular resolution submillimeter polarimeters such as POL-2 or HAWC+ which are capable of probing magnetic fields in high-density filaments will be helpful in deciphering the role of magnetic fields in the evolution from filaments to star-forming cores.
 

\section{Conclusions}\label{sec_summary}
We present JCMT POL-2 850 $\mu$m polarization observations of the DR21 filament.
With the Planck 850 $\mu$m dust polarization data, we were able to characterize the magnetic field structures from the surrounding ambient gas to the DR21 filament at scales from 10 pc to 0.1 pc. Our main results are the following:

\begin{enumerate}
\item 
The POL-2 data reveal ordered parsec-scale magnetic fields that are perpendicular to the DR21 main filament and parallel to the sub-filaments. 
The magnetic fields of the sub-filaments appear to smoothly connect to the magnetic fields of the main filament. 
The magnetic fields revealed in the Planck data are well aligned with those of the POL-2 data, indicating a smooth variation of magnetic fields from large to small scales. 
 
 \item
The comparison of the total and polarized flux of the POL-2 and Planck data indicates that the missing flux issue of the POL-2 DR21 observations is more severe in Stokes $I$ data than Stokes $Q$ and $U$ data. In addition, the large polarization fractions ($\gg$ 20\%) of POL-2 low-intensity data and the preferentially large polarization fractions of POL-2 data than SCUPOL data can be explained by the Stokes $I$ missing flux. 
 
\item
We find a power index $\alpha$ of 0.30--0.34 of the correlation between the polarization fractions and Stokes $I$ intensities of POL-2 data. The $\alpha$ value is consistent with those inferred from the POL-2 observations toward massive star-forming regions Orion B and NGC 6334 but shallower than the POL-2 observations toward less massive clouds, suggesting that the dust grain alignment efficiency of DR21 main filament is strongly influenced by the stellar radiation from the newborn stars. 

\item 
The analysis of the angular dispersion functions of dust polarization yields $B_{\rm pos}$ of 0.6--1.0 mG in the DR21 filament and $\sim$ 0.1 mG in the surrounding ambient gas.
The material is found to be magnetically supercritical in the filament and slightly subcritical to nearly critical in the ambient gas, consistent with the observed global infall motions of the DR21 filament. The sub-filaments might be the places where the transition from subcriticality to supercriticality occurs. 

\item 
The histogram of relative orientations between the density gradient and the magnetic field of the DR21 filament decreases with increasing $N(H)$ from no preferred alignment in the low-density ambient gas to mostly perpendicular in the high-density filament, in agreement with the HROs in other clouds.
Owing to the high angular resolution of POL-2, we are able to trace the HRO in the highest $N(H)$ regime to date.
A tentative positive slope of the HRO in the high-density DR21 filament is also found, as suggested from the complex magnetic field structures of the star-forming cores in the filament.

\end{enumerate} 

In summary, the analyses including the $B_{\rm pos}$, magnetic criticality, and histogram of relative orientations are all in good agreement with recent MHD simulations of a strongly magnetized medium, suggesting that magnetic fields play an important role in shaping the main filament and sub-filaments of the DR21 region.

\acknowledgments
The James Clerk Maxwell Telescope is operated by the East Asian Observatory on behalf of the National Astronomical Observatory of Japan, the Academia Sinica Institute of Astronomy and Astrophysics, the Korea Astronomy and Space Science Institute, and the Center for Astronomical Mega-Science. Additional funding support is provided by the Science and Technology Facilities Council of the United Kingdom and participating universities in the United Kingdom, Canada, and Ireland. Additional funds for the construction of SCUBA-2 and POL-2 were provided by the Canada Foundation for Innovation.
The authors wish to recognize and acknowledge the very significant cultural role and reverence that the summit of Mauna Kea has always had within the indigenous Hawaiian community.  We are most fortunate to have the opportunity to conduct observations from this mountain.

This work is supported by National Natural Science Foundation of China (NSFC) grant Nos.\ 11988101, U1931117, 11725313, and 12073061 and the CAS International Partnership Program of Chinese Academy of Sciences grant No.\ 114A11KYSB20160008. 
T.-C.\ C.\ is funded by Chinese Academy of Sciences Taiwan Young Talent Program Grant No.\ 2018TW2JB0002.
T.-C.\ C.\ and C.\ E.\ were supported by Special Funding for Advanced Users, budgeted and administrated by Center for Astronomical Mega-Science (CAMS), Chinese Academy of Sciences.
K.\ P.\ is a Royal Society University Research Fellow, supported by grant number URF$\backslash$R1$\backslash$211322.
K.Q.\ is partially supported by National Key R\&D Program of China No.\ 2022YFA1603100, and acknowledges the National Natural Science Foundation of China (NSFC) grant U1731237. 
S.P.L.\ acknowledges grants from the Ministry of Science and Technology of Taiwan 106-2119-M-007-021-MY3 and 109-2112-M-007-010-MY3. 
Y.D.\ acknowledges the support of JSPS KAKENHI grants 25247016 and 18H01250. 
Y.S.D.\ is supported by the National Key R\&D Program of China for grant No. 2022YFA1605000, and NSFC grants No. 12273051, 11933003.
D.\ J.\ is supported by the National Research Council of Canada and by a Natural Sciences and Engineering Research Council of Canada (NSERC) Discovery Grant.
P.\ M.\ K.\ is supported by the Ministry of Science and Technology (MoST) in Taiwan through grants 109-2112-M-001-022 and 110-2112-M-001-057.
C.\ W.\ L.\ is supported by the Basic Science Research Program through the National Research Foundation of Korea (NRF) funded by the Ministry of Education, Science and Technology (NRF-2019R1A2C1010851), and by the Korea Astronomy and Space Science Institute grant funded by the Korea government (MSIT; Project No.\ 2022-1-840-05).
T.\ H.\ is supported by the National Research Foundation of Korea (NRF) grant funded by the Korea government (MSIT) through the Mid-career Research Program (2019R1A2C1087045).
W.\ K.\ was supported by the NRF grant funded by the MSIT (2021R1F1A1061794).
C.\ E.\ acknowledges the financial support from grant RJF/2020/000071 as a part of Ramanujan Fellowship awarded by Science and Engineering Research Board (SERB), Department of Science and Technology (DST), Govt.\ of India.
F.\ P.\ acknowledges support from the Spanish State Research Agency (AEI) under grant number PID2019-105552RB-C43.
M.\ T.\ is supported by JSPS KAKENHI grant Nos.\ 18H05442, 15H02063, and 22000005. J.\ K.\ is supported by JSPS KAKENHI grant No.\ 19K14775.
L.\ F.\ and F.\ K.\ acknowledge the support by the MoST in Taiwan through grant 107-2119-M-001-031-MY3 and Academia Sinica through grant AS-IA-106-M03.
L.\ F.\ acknowledges the support by the MoST in Taiwan through grants 111-2811-M-005-007 and 109-2112-M-005-003-MY3.
C.\ L.\ H.\ H.\  acknowledges the support of the NAOJ Fellowship and JSPS KAKENHI grants 18K13586 and 20K14527.

\clearpage	

\appendix
\section{The Transformation of a Position Angle from Galactic Coordinate to Equatorial Coordinate}\label{app_a}
\begin{figure*}[h!]
\centering
\includegraphics[scale=1]{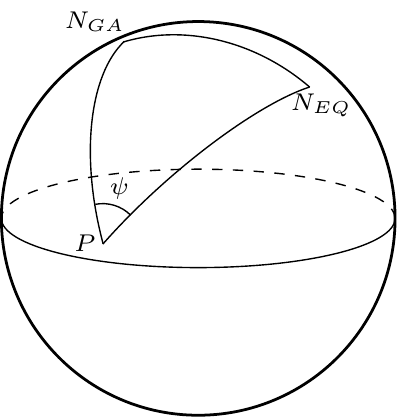}
\caption{Illustration of the the angle $\psi$ and the spherical triangle of $N_{GA}$, $N_{EQ}$, and $P$.} 
\label{fig_app}
\end{figure*}

In the IAU convention, the orientation of a position angle and a polarization angle is measured from North and positively towards East. At a position $P$ on the sky, the position angle measured in galactic coordinate ($PA_{GA}$) is different to the position angle measured in equatorial coordinate ($PA_{EQ}$) by the angle $\psi$ between the galactic North Pole ($N_{GA}$) and the equatorial North Pole ($N_{EQ}$) as 
\begin{equation}
PA_{EQ} = PA_{GA} - \psi.
\end{equation}
According to spherical trigonometry, $\psi$ can be derived with two sides and an opposite angle given (Figure \ref{fig_app}). That is, with the side $\overline{N_{GA} N_{EQ}} = b_{N_{GA}} - b_{N_{EQ}}$, the side $\overline{N_{GA} P} = b_{N_{GA}} - b_P$, and the angle $\angle N_{EQ}N_{GA}P = l_{N_{EQ}}-l_P$,
\begin{equation}
\tan(\psi) = \frac{\sin(b_{N_{GA}}-b_{N_{EQ}})\sin(l_{N_{EQ}}-l_P)}
{\sin(b_{N_{GA}} - b_P) \cos(b_{N_{GA}}-b_{N_{EQ}}) - \cos(b_{N_{GA}} - b_P)\sin(b_{N_{GA}}-b_{N_{EQ}})\cos(l_{N_{EQ}}-l_P)} ,
\end{equation}
where $b_{N_{GA}}, b_{N_{EQ}}, l_{N_{EQ}}, b_P$, and $l_P$ are the galactic latitudes and longitudes of $N_{GA}, N_{EQ}$, and $P$.
After substituting $b_{N_{GA}} = 90^\circ, b_{N_{EQ}}= 27.1^\circ, l_{N_{EQ}}= 122.9^\circ$  for epoch J2000 and some algebraic manipulations,
\begin{equation}
\psi = \arctan \left[ \frac{\cos(l_P-32.9^\circ)}{\cos b_P \cot 62.9^\circ - \sin b_P \sin(l_P-32.9^\circ)} \right].
\end{equation}

\end{document}